\documentclass[sigconf]{acmart}

\pdfoutput=1

\usepackage{endnotes}
\usepackage{wrapfig}
\usepackage{color}
\usepackage{subcaption}
\usepackage{booktabs}
\usepackage{listings}
\usepackage{arydshln}
\usepackage{array}
\usepackage{tabularx}
\usepackage{makecell}
\usepackage{xspace}
\usepackage{balance}
\usepackage{multirow}
\usepackage{bigstrut}
\usepackage[super]{nth}
\usepackage{tikz}
\usepackage{amsmath,amssymb}
\usepackage{stackengine}
\usepackage{enumitem}

\usepackage[utf8]{inputenc}
\microtypecontext{spacing=nonfrench}
\DeclareUnicodeCharacter{394}{\change}

\newcommand{\etc}{etc.}
\newcommand{\eg}{e.g., }
\newcommand{\ie}{i.e., }

\expandafter\def\expandafter\wrapfigure\expandafter{\wrapfigure{o}{2in}\large\raggedright}

\newcommand{\point}[1]{\par\smallskip\noindent\textbf{#1. }}

\newcolumntype{L}[1]{>{\raggedright\arraybackslash}p{#1}}

\newcommand{\CJq}{``cookie jar''\xspace}

\newcommand{\CCq}{``clean crawl''\xspace}
\newcommand{\JS}{JavaScript\xspace}

\newcommand{\RM}{``readermode''\xspace}

\newcommand{\ManuallyReviewedPaywallSites}{115\xspace}
\newcommand{\circumventSet}{32\xspace}
\newcommand{\costSubscription}{105\xspace}

\newcommand{\thirdPartyPaywallsPct}{25\%\xspace}
\newcommand{\NumSubscriptions}{10\xspace}
\newcommand{\dataFromExtensions}{147\xspace}
\newcommand{\paywallTpLibraries}{43\xspace}
\newcommand{\paywallSitesFromLib}{1,563\xspace}
\newcommand{\datasetSize}{1,710\xspace}
\newcommand{\paywallCountries}{61\xspace}

\newcommand{\obfuscatedArticles}{48.2\%\xspace}
\newcommand{\truncatedArticles}{44.5\%\xspace}
\newcommand{\redirectedArticles}{7.3\%\xspace}
\newcommand{\bounceRatePaywalled}{68.4\%\xspace}
\newcommand{\bounceRateNonPaywalled}{67.5\%\xspace}
\newcommand{\softPaywalls}{66.7\%\xspace}
\newcommand{\hardPaywalls}{15.7\%\xspace}
\newcommand{\freemiumPaywalls}{16.6\%\xspace}
\newcommand{\freeArticlesMedianAll}{3.5\xspace}
\newcommand{\freeArticlesMedianSoft}{4\xspace}
\newcommand{\backLinkDiff}{18.9$\times$\xspace}
\newcommand{\pageviewDiff}{13.42\%\xspace}
\newcommand{\spentTimeDiff}{$2.46\times$\xspace}

\newcommand{\paywallAdoptionUS}{18.75\%\xspace}
\newcommand{\paywallAdoptionAU}{12.69\%\xspace}
\newcommand{\ClassifierEvalFMeasure}{75\%\xspace}
\newcommand{\ClassifierEvalPrecision}{77\%\xspace}
\newcommand{\ClassifierEvalRecall}{77\%\xspace}
\newcommand{\ClassifierEvalROCAUC}{0.74\xspace}
\newcommand{\monthlySubs}{82.86\%\xspace}
\newcommand{\monthlyOnlySubs}{64.76\%\xspace}
\newcommand{\annualSubs}{35.23\%\xspace}
\newcommand{\annualOnlySubs}{17.14\%\xspace}

\lstdefinelanguage{JavaScript}{
  keywords={typeof, new, true, false, catch, function, return, null, catch, switch, var, if, in, while, do, else, case, break},
  keywordstyle=\color{blue}\bfseries,
  ndkeywords={class, export, boolean, throw, implements, import, this},
  ndkeywordstyle=\color{darkgray}\bfseries,
  identifierstyle=\color{black},
  sensitive=false,
  comment=[l]{//},
  morecomment=[s]{/*}{*/},
  commentstyle=\color{purple}\ttfamily,
  stringstyle=\color{red}\ttfamily,
  morestring=[b]',
  morestring=[b]"
}

\newcommand{\PaywallSiteTypeNews}{80.3\%\xspace}
\newcommand{\PaywallSiteTypeRest}{19.7\%\xspace}
\newcommand{\PaywallUsePopular}{8.54\%\xspace}
\newcommand{\PaywallUsePopularityMedian}{365,316\xspace}
\newcommand{\PaywallLibPiano}{23.5\%\xspace}
\newcommand{\PaywallLibTecnavia}{21.0\%\xspace}

\copyrightyear{2020}
\acmYear{2020}
\acmConference[WWW '20]{Proceedings of The Web Conference 2020}{April 20--24, 2020}{Taipei, Taiwan}
\acmBooktitle{Proceedings of The Web Conference 2020 (WWW '20), April 20--24, 2020, Taipei, Taiwan}
\acmPrice{}
\acmDOI{10.1145/3366423.3380217}
\acmISBN{978-1-4503-7023-3/20/04}



\begin{document}
\pagestyle{plain} 

\title{Keeping out the Masses: Understanding the Popularity and Implications of Internet Paywalls}

\author{Panagiotis Papadopoulos}
\affiliation{
	\institution{Brave Software}
}

\author{Peter Snyder}
\affiliation{
	\institution{Brave Software}
}

\author{Dimitrios Athanasakis}
\affiliation{
	\institution{Brave Software}
}

\author{Benjamin Livshits}
\affiliation{
	\institution{Brave Software\\Imperial College London}
}

\begin{abstract}
Funding the production of quality online content is a pressing problem for content producers.  The most common funding method, online advertising, is rife with well-known performance and privacy harms, and an intractable subject-agent conflict: many users do not want to see advertisements, depriving the site of needed funding.

Because of these negative aspects of advertisement-based funding, \emph{paywalls} are an increasingly popular alternative for websites.  This shift to a ``pay-for-access'' web is one that has potentially huge implications for the web and society.  Instead of a system where information (nominally) flows freely, paywalls create a web where high quality information is available to fewer and fewer people, leaving the rest of the web users with less information, that might be also less accurate and of lower quality. Despite the potential significance of a move from an ``advertising-but-open'' web to a ``paywalled'' web, we find this issue understudied. 

This work addresses this gap in our understanding by measuring how widely paywalls have been adopted, what kinds of sites use paywalls, and the distribution of policies enforced by paywalls. A partial list of our findings include that (i) paywall use has increased, and at an increasing rate ($2\times$ more paywalls every 6 months), (ii) paywall adoption differs by country (\eg \paywallAdoptionUS in US, \paywallAdoptionAU in Australia), (iii) paywall deployment significantly changes how users interact with the site (\eg higher bounce rates, less incoming links), (iv) the median cost of an annual paywall access is 108 USD \textit{per site}, and (v) paywalls are in general trivial to circumvent.

Finally, we present the design of a novel, automated system for detecting whether a site uses a paywall, through the combination of runtime browser instrumentation and repeated programmatic interactions with the site. We intend this classifier to augment future, longitudinal measurements of paywall use and behavior.
\end{abstract}

\maketitle

\section{Introduction} 
\label{Introduction} 

Publishers are increasingly moving away from ad-based models, because of the well-known failures~\cite{2018adcost} of ad-based internet funding models.  The most common adopted alternative is for sites to deploy ``paywalls''.
``Paywalls'' here are a broad term for monetization systems where visitors are charged subscription fees to access site content, sometimes after being able to sample a small amount of content for free.
The upsides of paywall systems are well understood (\ie they promise to enable the continued creation of high-quality content).  Less understood are the risks and larger implications of an increasingly ``walled'' web.  Possible risks include reducing societal access to news and information and the privacy harms of the increased user tracking needed to enforce paywalls.

This work aims to improve the understanding of the popularity, risks and benefits of paywalls online.  To introduce the topic, we first (i) describe why the web is increasingly moving away from ``open'' models to ``paywalled'' models, (ii) outline
why this transition is an important topic of study for the research community,
and then (iii) present the structure of the rest of the paper.

\subsection{The Move from Ads to Paywalls}
Digital advertising is the current dominant monetization method for web
publishers, and funds much of the web.  Publishers sell advertisements along
page content; middle parties buy these ad slots and fill them with images and
content provided by clients and ad-agencies.  This process is usually
programmatic, based of user's personal (\ie behavioral) data, and completed via
real-time programmatic auctions~\cite{rtbPrices17,pachilakis2019no}.

Web sites are increasingly unsatisfied from this ad-based funding system, for
many reasons.  First, the system is dominated by two parties, Google and
Facebook, who jointly harvest more than 70\% of global ad
revenues~\cite{adOligopoly,adDuopoly}, reducing the publisher's ``take'' for ad placements through market power. Second, ad-based funding systems suffer from significant and increasing rates of
fraud~\cite{adfraud,adfraud2,Liu:2014:DDC:2616448.2616455,Zarras:2014:DAM:2663716.2663719,malvertising},
depriving web sites of further funding.  Third, behavioral advertising
systems are increasingly incompatible with individual and legal privacy
demands~\cite{trackersWWW2016,vallina2016tracking,Leung:2016:YUA:2987443.2987456,DBLP:journals/corr/abs-1805-10505,razaghpanah2018apps}.
Last, users increasingly use ad blocking tools, for a variety of privacy, performance, and aesthetic reasons~\cite{adlergic,nithyanand2016adblocking},
further depriving publishers of revenue. As a result, ad revenues have decreased in recent years. Both big and small publishers are
coming up short on advertising revenue, even if they are long on visitors
traffic. Accounts of publisher-loss under ad-based funding models contain figures as high as~95\%~\cite{techdirtAdLoss}.

The difficulties of ad-based funding systems have pushed publishers to
alternative funding models, including
donations~\cite{wikipediaFunds1,wikipediaFunds2} or in-browser
crypto-mining~\cite{truthMiners2018}. The most common alternative though is ``paywalls'',
where users pay publishers directly to
access the content they create~\cite{drum2019} 
Figure~\ref{fig:screenshot} shows a representative example of a paywall system.

Paywalls so far have a mixed record as funding systems for publishers.
Publishers with large, loyal audiences and high-quality content tend to
be successful with this subscription strategy, with The New York
Times~\cite{nyTimesPaywall}, Wired~\cite{wiredPaywall}, The Financial Times~\cite{ftPaywall} and The Wall Street  Journal~\cite{wsjPaywall} as
prominent successful examples. The success of paywalls for smaller and more targeted sites (\eg local news), or sites with
less affluent audiences, is less clear.


It is important to note that the rapid growth of paywalls has drawn the attention of big tech companies like Google, Facebook and Apple, who have started building platforms to provide or support paywall
services~\cite{paywallsApple,paywallsGoogle,paywallGoogle2,paywallsFacebook}, in an effort to claim their share of the market.

\subsection{Understanding the State of Paywalls} 
Creating a sustainable system to fund news and related content is an important goal, and paywalls seem to be a promising (partial) solution to the problem. However, this move from ``open'' to ``walled'' business strategies brings significant, understudied risks. For example, paywalls (implicitly or otherwise) may impose a ``class system'' on the web~\cite{paywallsDemocracy1,paywallsDemocracy2}, potentially driving information-seeking visitors who cannot afford to pay for subscriptions to
badly-sourced, less-vetted, or even intentionally false (but free) new sources.

Despite the importance of the rise of paywalls to the web, it is surprising how little the topic has been studied by the research community.  Important open
questions include how popular paywall systems are, what policies paywalls
impose, how users are tracked for paywall enforcement, and whether paywalls are effective at protecting premium content.

\subsection{Contributions}
In this work, we aim to improve the understanding of paywall systems through the first systematic study of paywalls on widely-used web sites. This work makes the following contributions to the understanding of paywall systems on the web:

\begin{enumerate}[labelindent=1cm, leftmargin=0.5cm]
  \item A \textbf{novel system for programmatically determining if a site is using a paywall}, though the combination of multiple crowd-sourced data sets and tools.

  \item A \textbf{case study of how a popular paywall library operates},
    from how a publisher deploys it, how the paywall identifies
    users, to how the configured content access policy is enforced.

  \item A \textbf{large-scale measurement of paywall popularity}, including
    what kinds and what countries account for most paywall use, and how paywall use has changed over time.
    Example results include finding that paywall use has increased dramatically over time ($2\times$ more paywalls every 6 months) and that paywall adoption differs by country (\eg \paywallAdoptionUS in US, \paywallAdoptionAU in Australia) and industry.

  \item An \textbf{in-depth, large scale analysis of deployed paywall policies}, including subscription costs, how paywall adoption impacts the hosting website, how robust paywalls are to evasion, the mechanisms paywalls use to prevent users from viewing protected content, and the privacy implications of paywalls.

  \item A \textbf{classifier for deterring whether a site is using a paywall}
    for use on sites not considered by crowd-sourced resources, for future long
    term, web scale measurements of paywall adoption and behavior. 
\end{enumerate}

\begin{figure}[t]
\begin{subfigure}{.5\textwidth}
  \centering
  \includegraphics[width=0.8\linewidth]{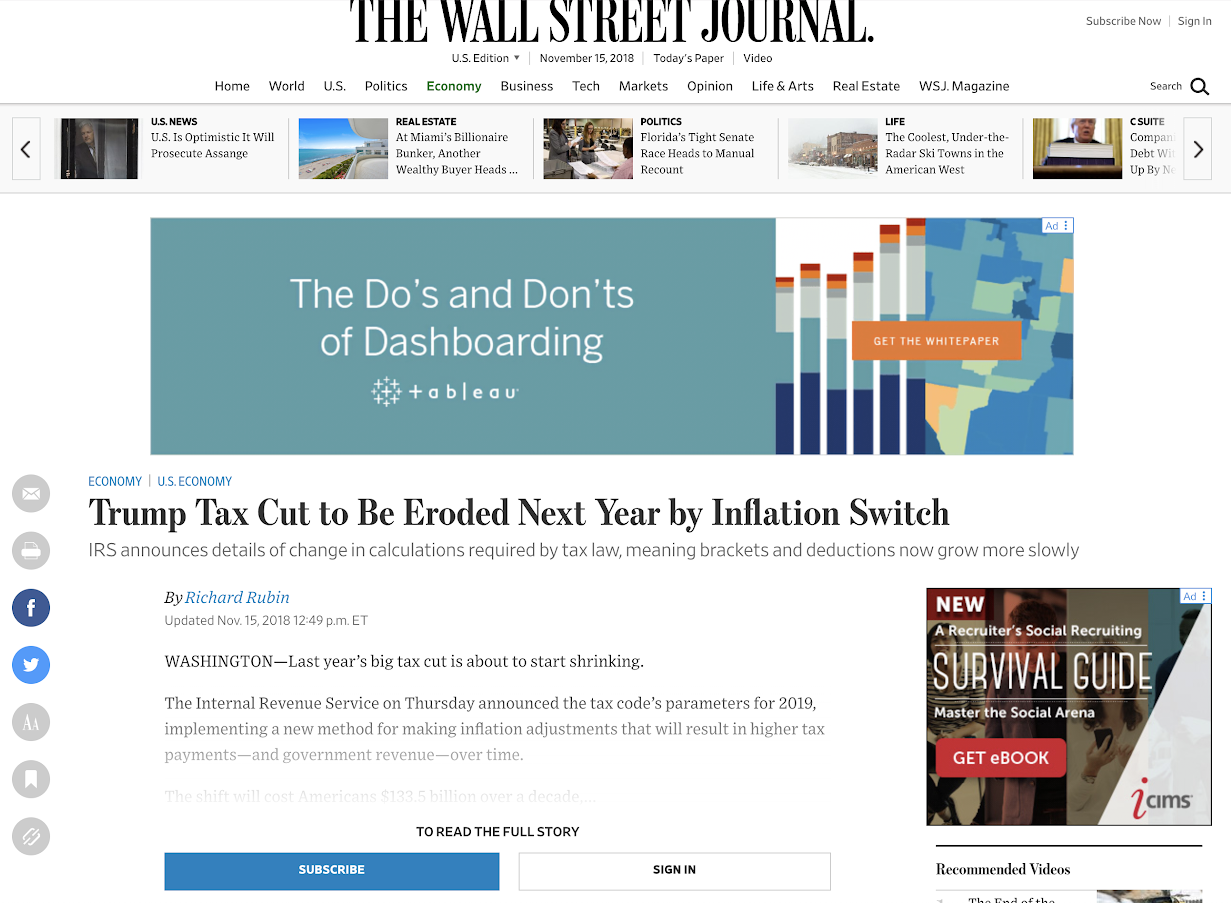}
  \caption{Truncated article in Wall Street Journal.}
  \label{fig:screenshot1}
\end{subfigure}%
\hfill
\begin{subfigure}{.5\textwidth}
  \centering
  \vspace{0.3cm}
  \includegraphics[width=0.8\linewidth]{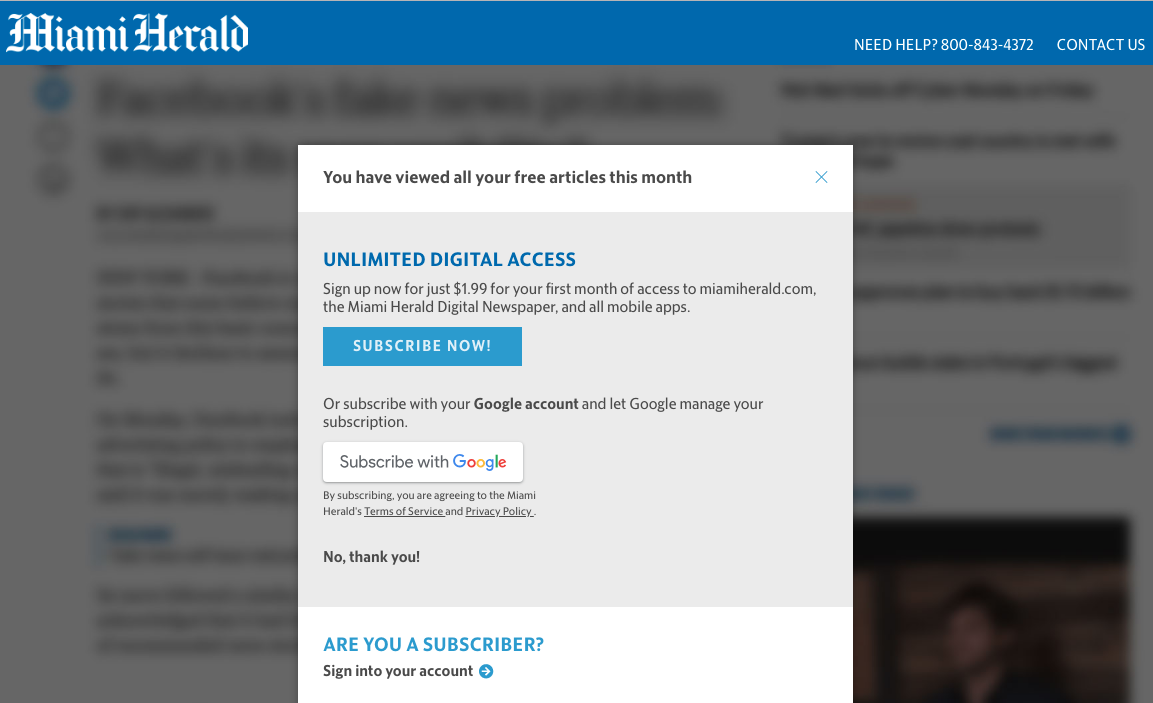}
  \caption{Obscured article in Miami Herald.}
  \label{fig:screenshot2}
\end{subfigure}\vspace{-0.2cm}
\caption{Examples of raised paywalls in major news sites. Paywalls may be enforced in different ways to deny access to articles to non-subscribed users.}
\label{fig:screenshot}\vspace{-0.4cm}
\end{figure}
\section{Background} 
\label{Background}

Paywalls are an increasingly popular monetization strategy for web sites, as publishers attempt to become less dependent on advertising. Figure~\ref{fig:screenshot} shows a typical paywall, where a publisher is blocking access to content until the user pays a fee. To enforce access control, paywalls track the engagement of the user with the publisher content: \ie how much time they spend on a web site, how many articles they have read, how many times a user has visited the website. 

\subsection{Types of Paywalls}
We group paywalls into two categories, based on how restrictive they are: (i) \emph{hard paywalls}, where users cannot gain access the site without first purchasing a subscription (\eg monthly or annual subscriptions) and (ii) \emph{soft paywalls} that allow limited, free-of-charge viewing for a specific amount of time or number of visit (\eg~5 free articles per month per user). 

\point{Hard Paywalls} 
Hard paywalls require subscriptions before visitors can access content (\eg Financial Times requires a subscription before the user can read any article). Such a strategy runs the risk of deterring users and thereby diminishing the publisher's influence over all. As reported in the press~\cite{declinedTraffic}, The Times experienced a~90\% drop in  traffic after introducing a hard paywall.

\point{Soft or Metered Paywalls} 
Soft (or metered) paywalls limit the number of articles a viewer can read before requiring a paid subscription. Soft paywalls use the free articles as a strategy to entice users to subscribe. Soft paywalls require some method (often a JavaScript snippet on the user-side) for measuring either the number of articles a user has accessed, or the time a user spends in browsing the website's articles.

\begin{figure}[tb]
    \centering
    \vspace{-.9cm} 
    \includegraphics[width=1.18\linewidth]{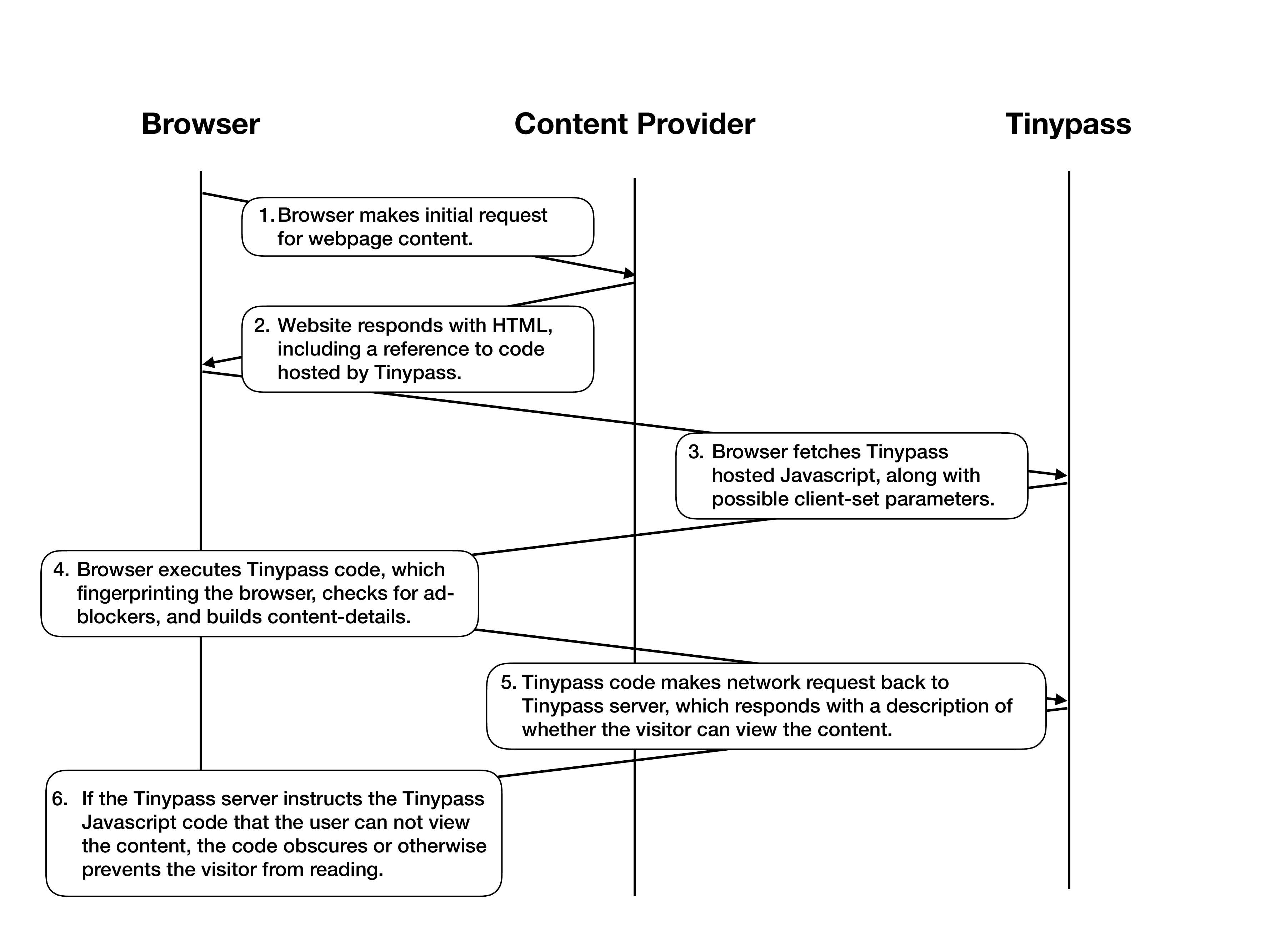}\vspace{-.5cm}
    \caption{High level overview of the core functionality of a paywalled website powered by Tinypass.}\vspace{-.55cm}
    \label{fig:tinypassOverview}
\end{figure}

As with hard paywalls, a publisher's web traffic can also be affected by the installation of soft paywalls (\eg traffic to the New York Times declined by~5\% to~15\% one month after the installation of its soft paywall~\cite{NYTdecreasedTraffic1,NYTdecreasedTraffic2}). Overall, though, fewer users are discouraged by soft paywalls. Prior studies~\cite{hardVsSoft}  have found, on average, retention rates for publishers with soft paywalls reaching~58.5\%, compared to only~15--20\% for publishers with hard paywall.
\section{Paywall Case Study}
\label{sec:casestudy}

This section provides a detailed case study of a popular third-party paywall system.  We provide this case study (i) to introduce the reader to how paywalls work, and (ii) to document the kinds of privacy-affecting behaviors paywalls often rely on to impose their policies.
We select Piano's Tinypass paywall-as-a-service product~\cite{tinypass} for our case study for several reasons.  First, it is one of the most popular third-party paywall providers (Tinypass owns~38.2\% of the market, as measured in Figure~\ref{fig:thirdPartyLibs_popularity}), so understanding how this system works provides a good understanding of the kinds of paywall code users are likely to experience. And second, Tinypass can be deployed as a configurable, paywall-as-a-service, allowing publishers (blogs, news sites, magazines, \etc) to impose a variety of paywall policies, both hard and soft.

\subsection{Tinypass: The protocol}
At some point prior to the user's visit, a site owner creates an account at Tinypass, where they describe the subscription policies they wish to enforce.  Tinypass generates the keys and identifiers used to enforce the paywall and track visitors. 
Once a site owner installs Tinypass on their site, the paywall works in the following six stages, with numbers corresponding to Figure~\ref{fig:tinypassOverview}:

\point{Step one} 
The user's browser makes a request to a website where the site owner has installed Tinypass.

\point{Step two} 
The website responds with the HTML of their page, including a reference to the Tinypass \JS library, hosted on Tinypass's servers. The content provider's response may also include optional, customized parameters that allow Tinypass to integrate with other services, like Facebook and Google Analytics. At the time of this writing, Tinypass's code is hosted at \url{https://code.tinypass.com/tinypass.js}.

\point{Step three} 
The referenced JavaScript causes the browser to request code from Tinypass's server, which responds with a bootstrapping system, providing basic routines for fetching the main implementation code, helper libraries, and utilities for rate limiting and fingerprinting.  Depending on the particular deployment, minified versions of this code also includes common utilities like CommonJS-style dependency tools or cryptography libraries.

\point{Step four} 
The browser executes the complete Tinypass library, and the full (post-bootstrap) Tinypass library performs a number of privacy-relevant checks. First, Tinypass attempts to determine if a site visitor is actually an automated browser (\eg Puppeteer, WebDriver client). Tinypass attempts to determine if the user has an ad-blocker installed. Interestingly, Tinypass not only detects if the user currently has an ad-blocker installed, but also if the visitor has changed their ad-blocker usage (\eg the user had an ad-blocker installed on a previous visit but no longer does, or vice versa).
    
\begin{figure}[t]
{\footnotesize
\begin{lstlisting}[caption=Excerpt of Tinypass's fingerprinting JavaScript.,label={lst:fingerprint}]
var _getFingerprint = function () {
    if (fingerprint) {
        return fingerprint;
    }
    var fingerprint_raw = _getLocality();
    fingerprint_raw += _getBrowserPlugin();
    fingerprint_raw += _getInstalledFonts();
    fingerprint_raw += _getScreen();
    fingerprint_raw += _getUserAgent();
    fingerprint_raw += _getBrowserObjects();
    fingerprint = murmurhash3.x64hash128(fingerprint_raw);
    util.debug("Current browser fingerprint is: " + fingerprint);
    return fingerprint;
};
\end{lstlisting}\vspace{-0.7cm}
}
\end{figure}
    
Tinypass then generates a user fingerprint, implemented with the code hosted at \url{https://cdn.tinypass.com/api/libs/fingerprint.js}.  The Tinypass fingerprinting library (shown in part in Listing~\ref{lst:fingerprint}) hashes together a number of commonly known semi-unique identifiers (installed plugins, preferred language, installed fonts, screen position, user agent, \etc) to build a unique identifier, hashed together using the MurmurHash3 hash algorithm~\cite{murmur3}). The result is an identifier that is consistent across cookie-clears, and so can re-identify users attempting some evasion techniques. Tinypass also reads, if available, a first-party cookie the library also uses to identify users.  When available, this cookie is used in place of the above fingerprint, to track how much content the user has visited.

\begin{figure}[tb]
{\footnotesize
\begin{lstlisting}[caption=Excerpt of returned Tinypass end point data (meter is Tinypass's terminology for a counter describing how much more non-paywalled content a user can view).,label={lst:metrics}
]
 ...
    "trackingId": "{jcx}H4sIAAAAAAAAAI2QW2vCQBCF_8s...",
    "splitTests": [],
    "currentMeterName": "DefaultMeter",
    "activeMeters": [
        {
            "meterName": "DefaultMeter",
            "views": 0,
            "viewsLeft": 4,
            "maxViews": 4,
            "totalViews": 0
        }
    ],
    ...
\end{lstlisting}\vspace{-0.5cm}
}
\end{figure}

\point{Step five} 
Next, the Tinypass library gathers the above information, combines it with information about the page, derived fingerprinting values, the date, and other similar data, and \texttt{POST}s them to a Tinypass endpoint~\footnote{\url{https://experience.tinypass.com/xbuilder/experience/execute?aid=*}}, which records information about the page view. The server then returns a JSON string describing a variety of information about the page view, and excerpt of which is presented in Listing~\ref{lst:metrics}.  This JSON string includes a wide variety of both user-facing and program-effecting values, including how many more pages the user is able to visit before the paywall is triggered, possibly new identifiers to rotate on the browsing session, whether the user has logged in and is known to Tinypass (e.g. the user logged in on a different domain owned by the same publisher). 
    
\point{Step six} 
Finally, the Tinypass code running on the browser enforces the described paywall policy. The code, client-side, uses the response data to decide how to respond to the page view, possibly by  obscuring page content or presenting a subscription offer dialog (by default, Tinypass offers pre-made-but-configurable modal and ``inline'' dialogues the website can check from). In the pages we observed, Tinypass only enforced subscription requirements (\ie preventing users from viewing content) after the above check was completed.  A side effect of this implementation decisions is that Tinypass's restrictions can be circumvented by simply blocking the Tinypass library (see Section~\ref{sec:circumvention}).
\section{Current Paywall Deployments}
\label{sec:results}
In this section, we present a large-scale measurement of paywall deployments on the web. The measurements presented give a broad assessment of how often paywalls are used (by country and by industry). We then present a variety of measurements of how deployed paywalls operate, including the access policies they enforce, their enforcement mechanisms, and how robust these paywalls are to circumvention.  The section begins with a description of how we gathered there relevant datasets for measurements, and then proceeds in the above described order.

\begin{figure}[t]
        \centering
        \vspace{-0.4cm}
    \includegraphics[width=.7\linewidth]{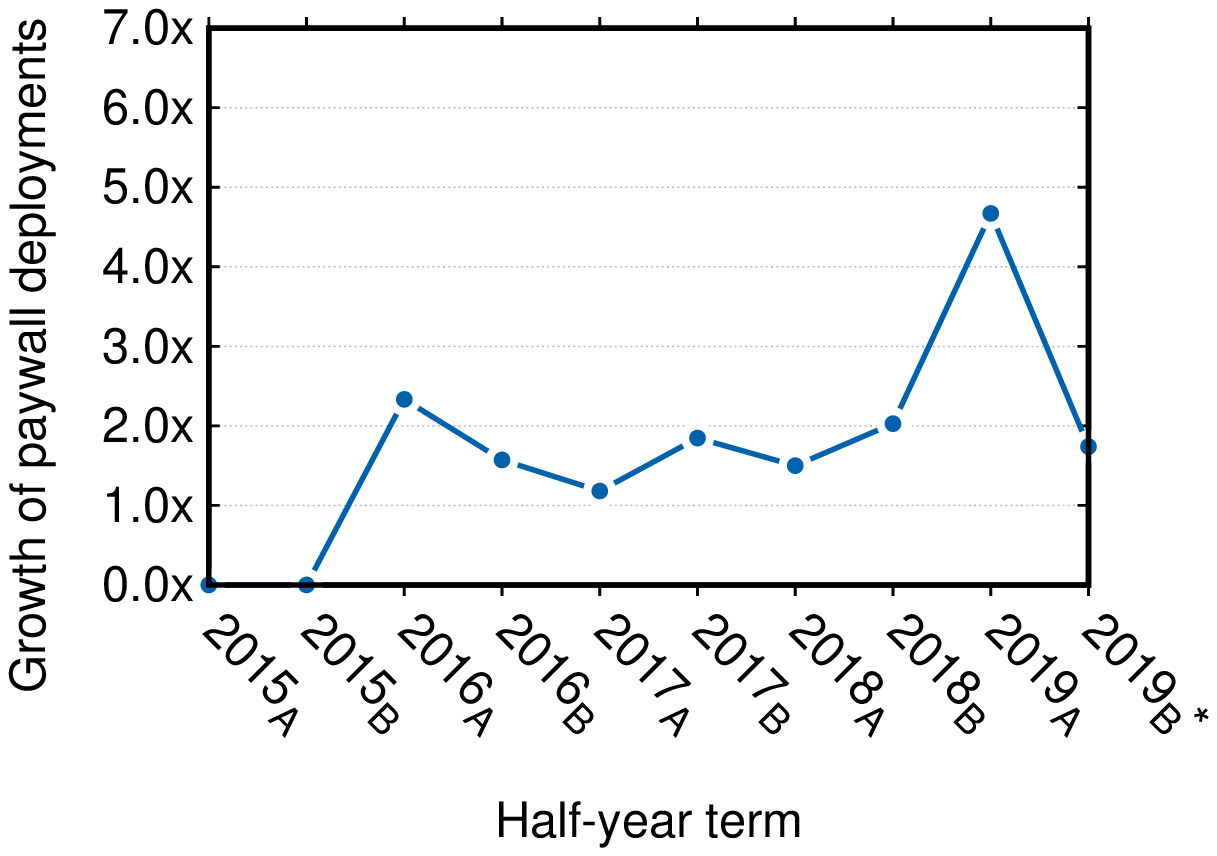}\vspace{-.2cm}
    \caption{Growth of paywall deployments per 6 months. Note that the y-axis
    depicts the growth-rate, and not absolute numbers.}\vspace{-.2cm}
    \label{fig:popularity_growth}
\end{figure}

\begin{figure}[b]
{\small
\centering
    \begin{tabular}{lr}
      \toprule 
            \textbf{Data} & \textbf{Volume} \\
        \midrule
            Paywalled websites from bypassing extensions & \dataFromExtensions \\
            Third-party paywall libraries & \paywallTpLibraries  \\
            Unique paywalled sites & \datasetSize \\
            Countries the paywalled sites originate from & \paywallCountries \\
        \bottomrule
    \end{tabular}
    }
    \vspace{-0.3cm}
    \caption{Summary of our crowdsourced dataset labeling which websites use
    paywalls.}
    \vspace{-0.3cm}
  \label{tbl:dataCollected}
\end{figure}

\subsection{Dataset}
\label{sec:dataset}
To conduct the measurements described in this section, we built
an oracle to determine whether a web site uses a paywall. While seemingly
a simple question, the diversity of paywall libraries, enforcement
mechanisms, access policies and varying verbiage makes this a difficult
question to answer without significant human intervention.
To solve this problem, we draw on two existing crowd-sourced datasets:  

\begin{figure*}[t]
    \centering
    \begin{minipage}{0.32\linewidth} 
    \centering\vspace{-.5cm}
    \includegraphics[width=1.1\linewidth]{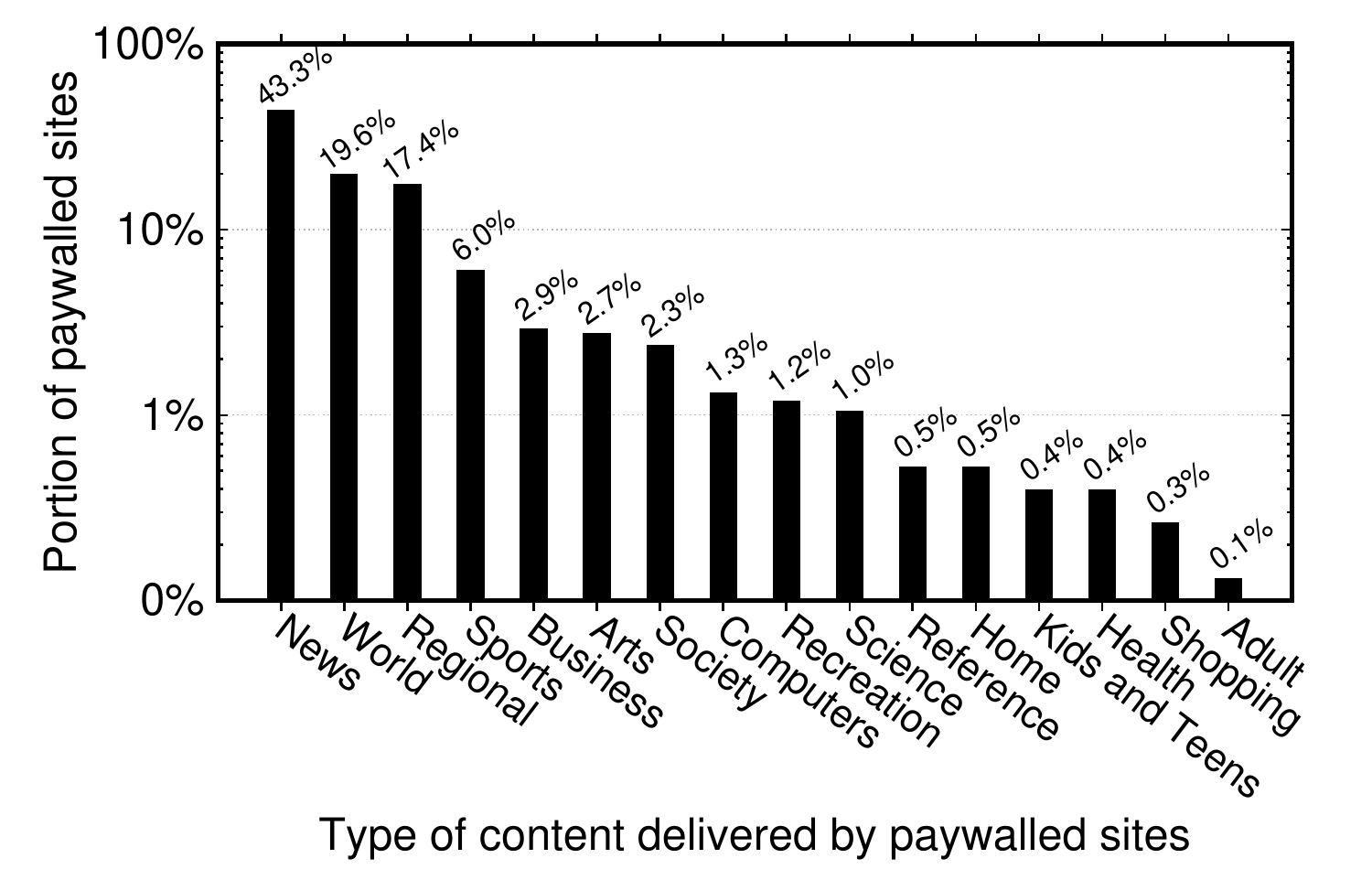}\vspace{-.3cm}
    \caption{Type of industry or type of content, paywalled websites deliver.}
    \label{fig:alexaTopic}
    \end{minipage}
    \hfill
    \begin{minipage}{0.32\linewidth} 
        \centering
        \vspace{.2cm}
    \includegraphics[width=1\linewidth]{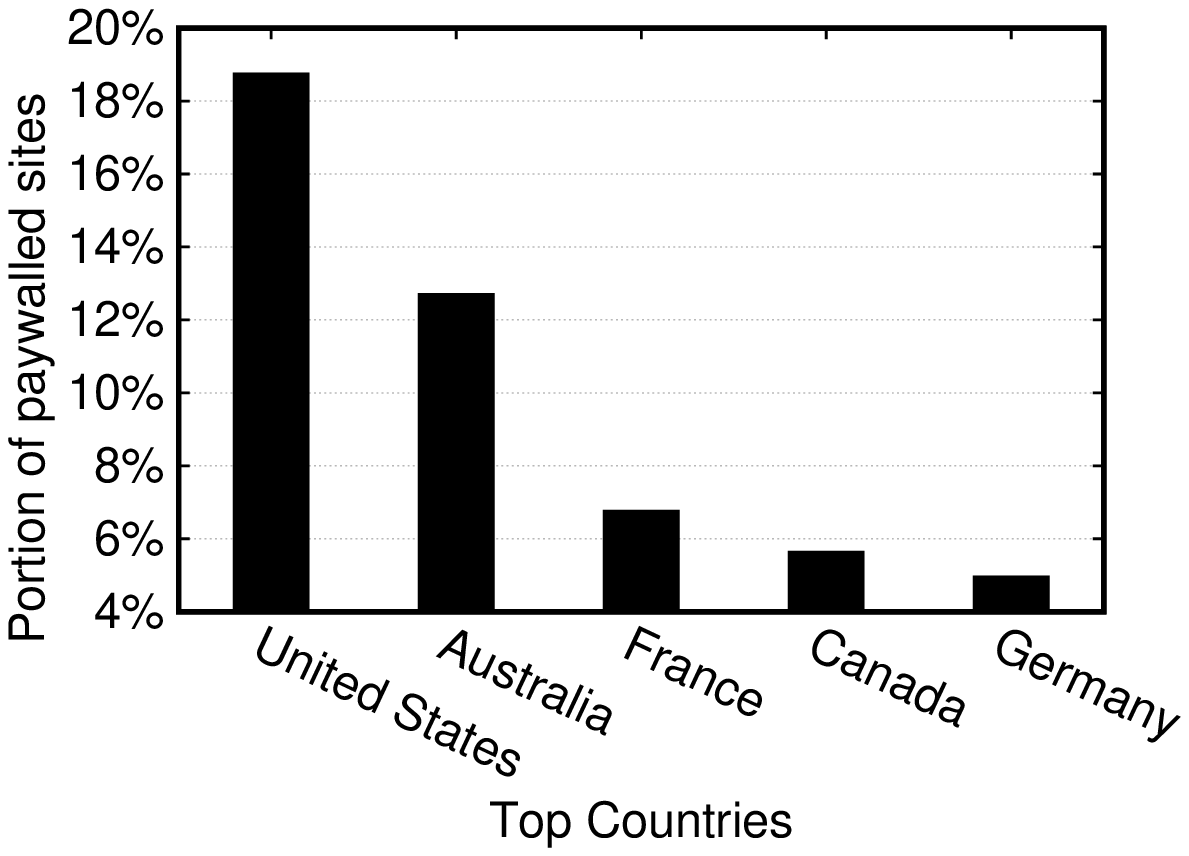}\vspace{-.3cm}
    \caption{Portion of news sites using paywalls per country. Paywall adoption reaches \paywallAdoptionUS in US and \paywallAdoptionAU in Australia.}
    \label{fig:popularityPerCountry}
    \end{minipage}
    \hfill
    \begin{minipage}{0.32\linewidth} 
        \centering\vspace{-0.3cm}
    \includegraphics[width=1\linewidth]{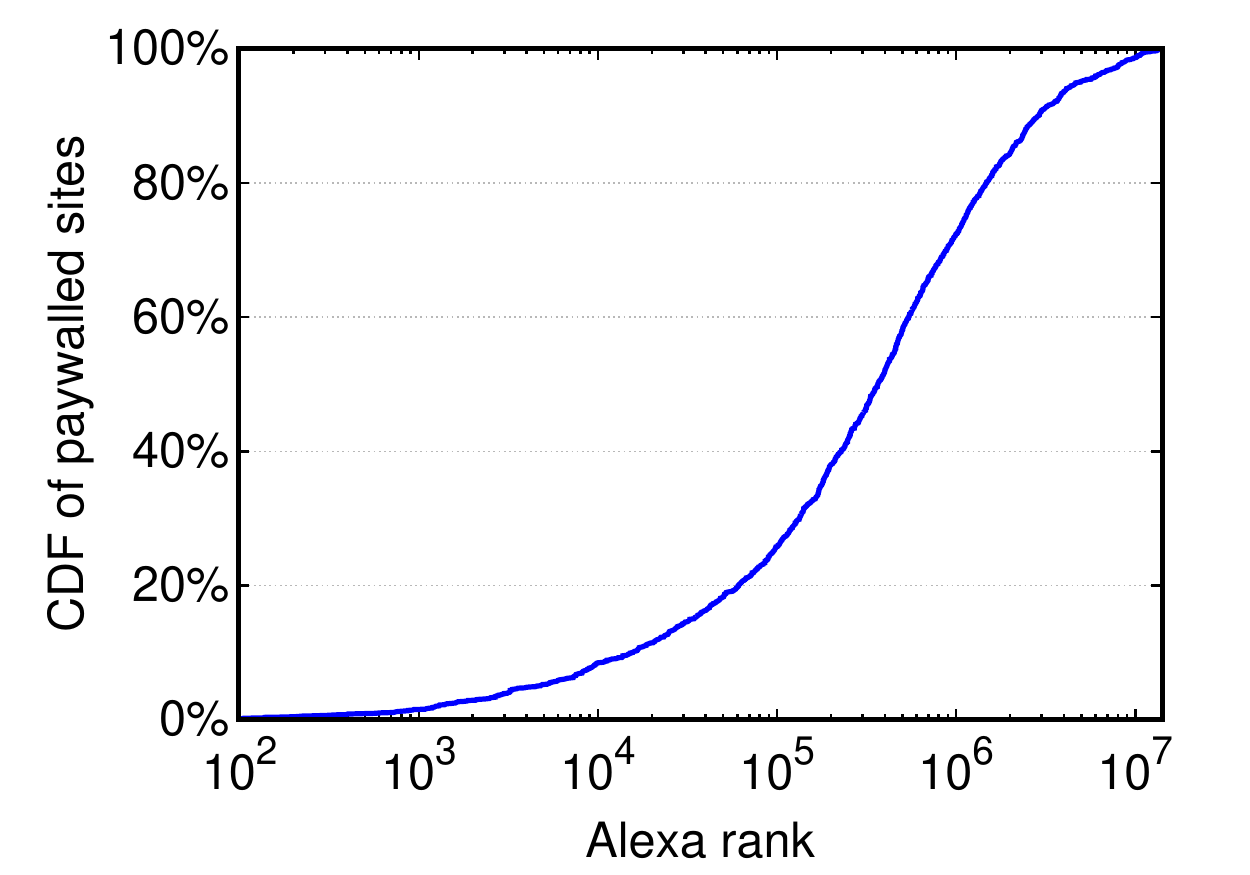}\vspace{-0.2cm}
    \caption{Distribution of the popularity of each paywalled website in our
    dataset based on the Alexa ranking.}
    \label{fig:alexaRank}
    \end{minipage}
    \vspace{-.4cm}
\end{figure*}

\point{A) Extensions}
First, we extract rules from several popular browser extensions~\cite{bypassExt1,bypassExt2,bypassExt3,bypassExt4} designed to help users circumvent paywalls.  By examining the source code of these extensions, we are able to (directly or indirectly) identify \dataFromExtensions{} websites that the tools' authors and maintainers label as using paywalls.

\point{B) Filter lists}
Second, we use a popular, crowd-maintained filter list that identifies
third-party paywall libraries~\cite{fanboyPaywalls} so that they can
be blocked with common filter-list consuming tools (\eg AdBlock Plus, uBlock Origin). This list includes filter rules for blocking resources related to a variety of internet ``annoyances''; we extract the subset of the list specifically targeting paywalls. This gives us a list of~\paywallTpLibraries{} third-party paywall libraries. We query for each entry of this paywall libraries list in two existing, current web crawl archives (\ie HTTPArchive~\cite{wayback} and PublicWWW~\cite{publicWWW}). We found \paywallSitesFromLib{} sites using one of these paywall libraries and we labeled them as ``paywalled''. 

We combine the above two approaches (\ie paywalled domains labeled by browser
extensions and paywalled sites including third-party paywall libraries)
to identify \datasetSize{} unique paywall-using domains, from
\paywallCountries{} countries.  This dataset, summarized in Figure~\ref{tbl:dataCollected}, comprises our complete dataset of paywalled sites used as an oracle in this section, and we provide open-sourced\footnote{\url{https://gist.github.com/panpap/68af1c99b49366dfce4044a354f6e1b8}}.

\subsection{Paywall Popularity}
We start by measuring how popular paywalls are, across several dimensions.
We use the domains identified in Section~\ref{sec:results} as the set of paywalled sites, and all other sites on the web as not paywalled. 

\subsubsection{Increase in Paywall Use}
\label{sec:growth}
We first measure whether paywall use has increased over time.  We find sites
in our dataset started using paywalls in~2015, with overall paywall use strictly
increasing since. Paywall use has increased at a rate between~120\% and~230\%
every six months since~2015 until recently. In the first six months of~2019,
paywall use quadrupled, and has grown by a further~\textit{180\% } during the
first two months in the second half of~2019. These measures are summarized
in Figure~\ref{fig:popularity_growth}.

We measure paywall growth over time by applying our paywall oracle
(described in Section~\ref{sec:dataset}) to archived versions of the same
sites in the Wayback Machine web archive~\cite{wayback}.  We use these
archived versions of each site to approximate date each site adopted a paywall.
The precise methodology is as follows:

\begin{enumerate}[leftmargin=0.5cm,topsep=0pt]
  \item We build the set of paywall library related URLs and domains using the technique described in Section~\ref{sec:dataset}.
  \item We fetch the most recent archive of each paywalled website in our dataset from the Wayback machine and check whether that historical version is using a paywall.
  \item If we observe the site using a paywall, we fetch the next-most-recent version of the site from the Wayback machine (\eg we move back one recording in time) and re-check.
  \item We continue this process until we encounter a version of the web site that no longer is using a paywall.  Once we encounter a non-paywalled version of the site, we note the date that version of the site and record it as when the site began using a paywall. \vspace{-0.1cm}
\end{enumerate}

\point{Limitations} 
We note two limitations of the above approach, and why we do not believe they significantly impact our findings. It is possible that earlier versions of sites used different types and providers of paywalls than current sites, and so our paywall detection oracle may be missing historical paywall use. While possible, we do not think this limitation significantly impacts the results for two reasons:  (i) prior research~\cite{vastel2018filters} has found that filter lists (like the ones we use for paywall library detection) rarely delete rules, and so that paywall-targeting filter lists would identify both current and historical paywalls.
What is more, (ii) we manually evaluated a random sample of commits from the git history of the paywall-targeting portion of the filter list and we found no rule deletions. This gives us further confidence, though not certainty, that filter rules that would identify paywalls on previous versions of paywalled sites have not been removed.

\begin{figure*}[t]
    \centering
    \begin{minipage}{0.32\linewidth}   
        \centering
        \includegraphics[width=1.05\linewidth]{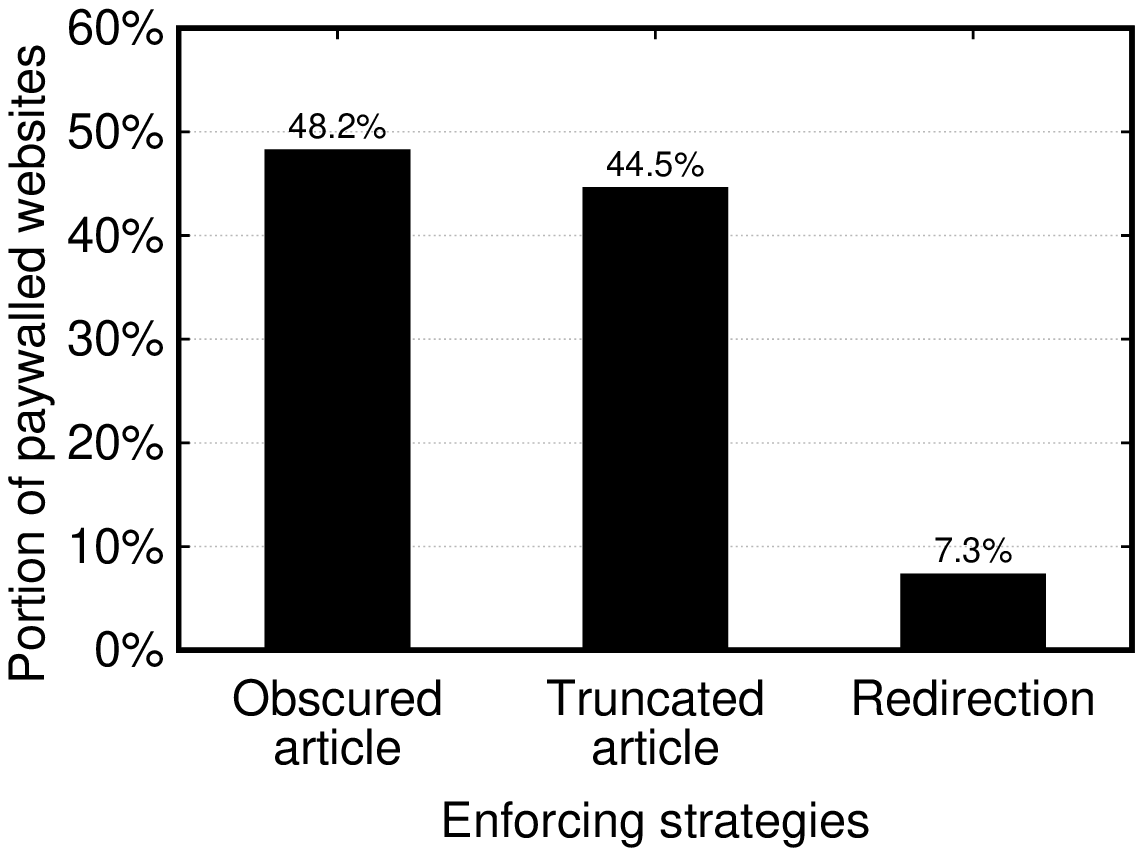}
        \caption{Popularity of the different paywall enforcing policies. Most of the publishers prefer to obfuscate (\obfuscatedArticles) or truncate (\truncatedArticles) the article the user has not yet access to.}
        \label{fig:enforcingStrategies}
        \end{minipage}
    \hfill
    \begin{minipage}{0.32\linewidth}   
        \centering\vspace{0.4cm}
        \includegraphics[width=1.05\linewidth]{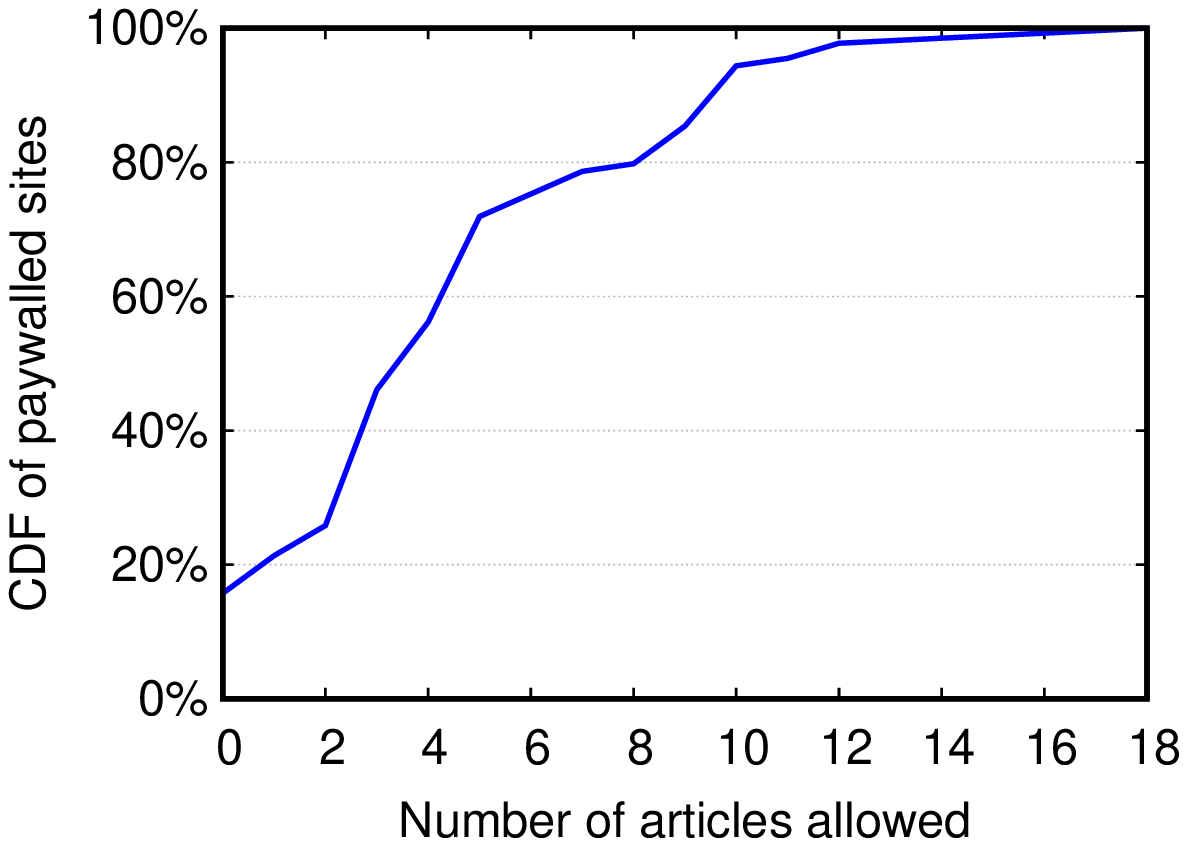}
        \vspace{-0.4cm}
        \caption{Distribution of the number of free articles allowed per user. The median paywalled website allows \freeArticlesMedianAll articles, the median soft-paywalled website allows \freeArticlesMedianSoft articles and all hard-paywalled do not allow \emph{any} free article.}
        \label{fig:freeArticles}
    \end{minipage}
    \hfill
    \begin{minipage}{0.32\linewidth} 
        \centering\vspace{-0.6cm}
    \includegraphics[width=1.05\linewidth]{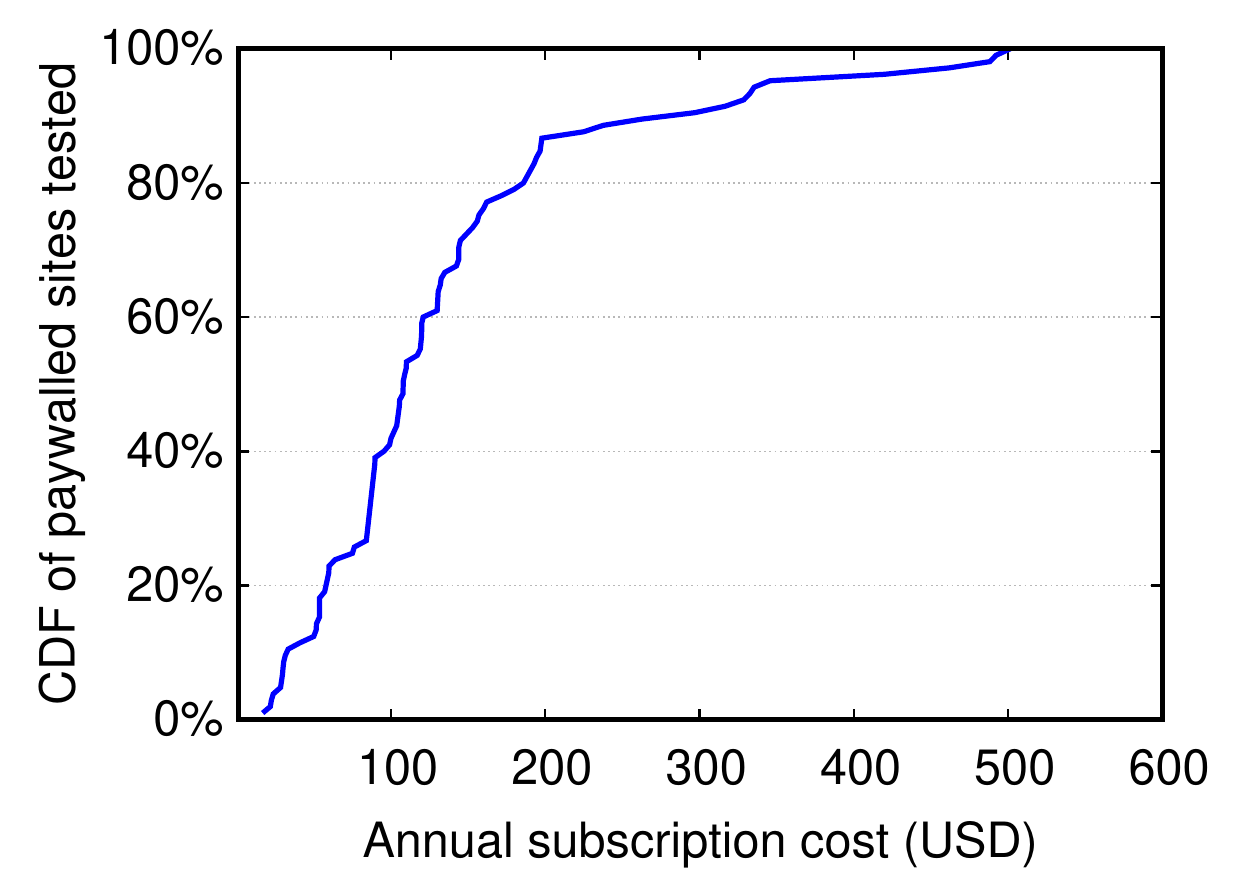}
    \vspace{-0.5cm}
    \caption{Cumulative distribution of the subscription cost per website for a 12-month content access.}
    \label{fig:cost_subscr_cdf}
    \end{minipage}
 \end{figure*}

The second possible limitation is that our approach might miss sites that
used to have paywalls, but no longer do. We believe such cases to be rare.
We observed no instances of sites using paywalls, removing the paywalls,
and then re-establishing it.  This suggests (though does not prove)
that sites do not commonly abandon paywall strategies once they have adopted them.

\subsubsection{Paywall Use by Site Type}
We measure what types of content paywalled sites provide.
We find that most~(\PaywallSiteTypeNews{}) paywalled sites provide some form
of news content, whether targeted at the local, regional, or world-level.
Figure~\ref{fig:alexaTopic} provides summary of this measurement.
For this measurement, we use the sites identified as using paywalls from
Section~\ref{sec:dataset} with information available from the Alexa Top Sites
service.  The Alexa Top Sites classifies domains into one of~17 different
classes (\ie news, sports, business, arts, society). Three categories
describe news content, though at different levels of focus (\eg ``World'',
``Regional'' or, generically, ``News'').  We group these together for our
measurements, since they are thematically very similar.  The remaining 14
categories account for just \PaywallSiteTypeRest{} of paywalled sites.

\subsubsection{Paywall Use by Country}
Next, we measure which countries have the highest rates of paywall use.  Because news sites account for most paywall use, we focus this measurement on news sites.  We find that US news sites have been the quickest to move to paywalls, followed by Australia, France, Canada and Germany.  Figure~\ref{fig:popularityPerCountry} summarizes our findings. Since our oracle does not identify all websites with paywalls, Figure~\ref{fig:popularityPerCountry} presents only the lower bound of the existing paywalled sites. 

We measure rates of paywall use by country by first retrieving the Alexa the Top~10,000 websites per country.  We filter the list and remove all non-news sites.  Then, we calculate the percentage of paywall-using news sites, as a fraction of all news sites, per country. We find that \paywallAdoptionUS of US news sites use paywalls, \paywallAdoptionAU of Australian new sites, and less than ~7\% in all other countries.

\subsubsection{Paywall Use by Popularity}
Next, we measure whether there is any clear relationship between paywall deployment and the popularity of a website. We did not observe any such relationship. We
anticipate that most paywalled sites would be popular (as measured by Alexa Top Sites), as a successful paywall would require a significant number of
subscribers, which in turn would require a significant amount of baseline
visitors. Instead, we find that only \PaywallUsePopular{} of paywall-using sites
are among the~10,000 most popular sites on the web.  The median paywall-using
site is ranked \PaywallUsePopularityMedian{}.  The full distribution
of the popularity of paywall using sites is presented in Figure~\ref{fig:alexaRank}.

\begin{figure}[t]
    \centering
    \vspace{-0.3cm}
    \includegraphics[width=0.9\linewidth]{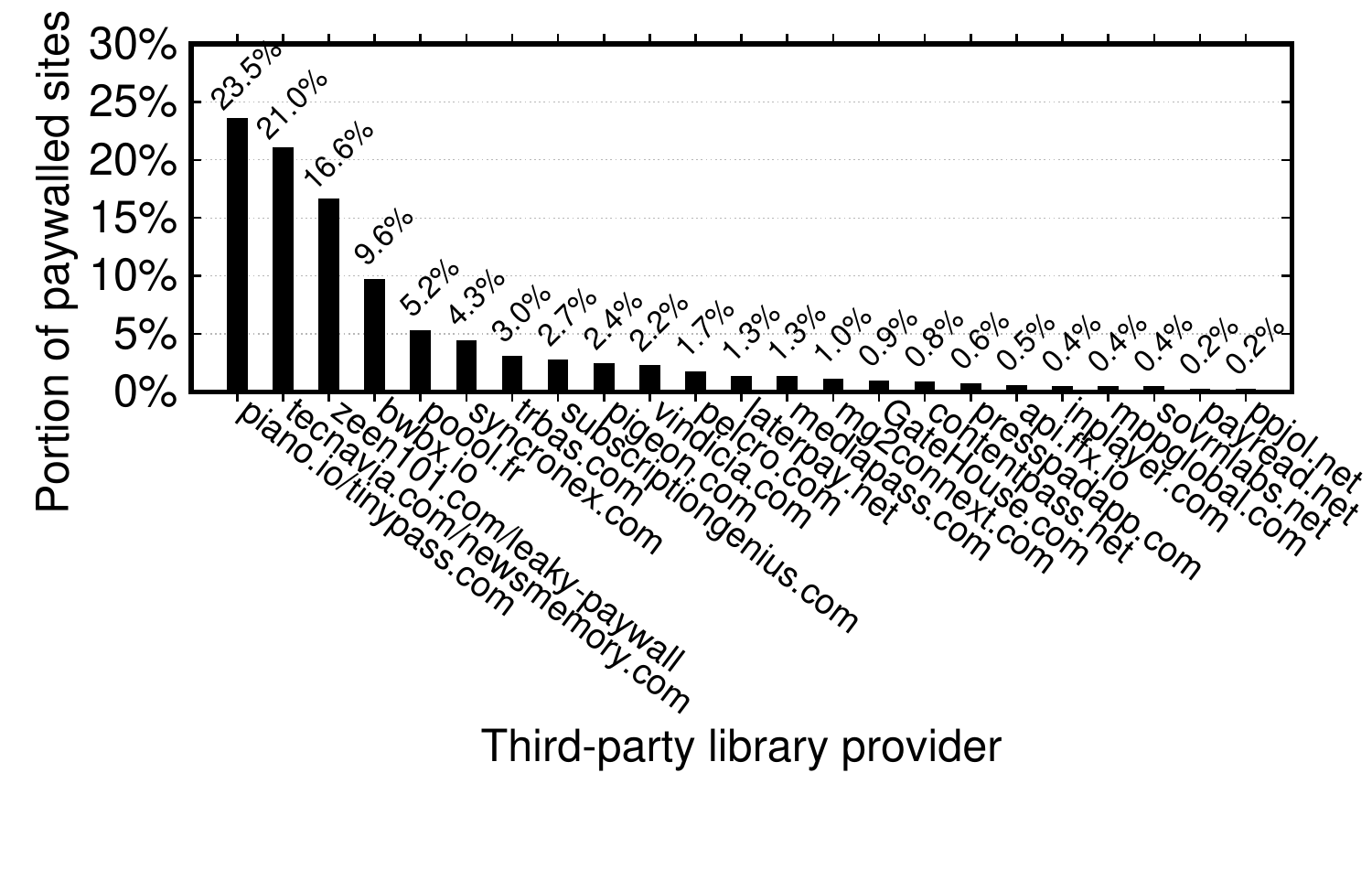}
    \vspace{-0.7cm}
    \caption{Popularity of third-party paywall libraries in our dataset. A
    small number of paywall implementations account for the majority of
    third-party paywall deployments.}
    \label{fig:thirdPartyLibs_popularity}\vspace{-0.2cm}
\end{figure}

\subsection{Paywall Libraries}
\label{sec:3plibs}
A significant number of sites rely on third-parties for their paywall
implementations. These third-parties sell ``paywall-as-a-service'' products,
where publishers pay fee to have the third-party manage and
enforce the paywall on the publisher's site.  We observe
that a small number of paywall providers account for the vast majority
of paywall deployments, with Piano and Tecnavia being the most popular
paywall providers (\PaywallLibPiano{} and \PaywallLibTecnavia{} market
share, respectively).  The full distribution of third-party paywall
market share is depicted in Figure~\ref{fig:thirdPartyLibs_popularity}.

This consolidation of paywall implementation and enforcement is significant,
for a variety of reasons.  First, market consolidation may effect the amount of
income content-makers can receive for their content (popular third-party
paywall providers receive 10-15\% of each sold subscription).  Second, provider
consolidation may make large scale circumvention easier, as circumventors
need to target a smaller number of systems (see Section~\ref{sec:circumvention}).  Third, a small number of paywall providers
tracking users across a large number of websites has clear privacy implications
(see Section~\ref{sec:privacy}).

We measure the popularity and consolidation of third-party paywall libraries 
by crawling each paywalled site in our dataset and observing which resources
from known paywall providers were fetched.  We find that at least
\thirdPartyPaywallsPct of paywalled websites outsource their paywall
functionality to third-parties. The distribution of third party paywall
use follows a rough power-law distribution.

\subsection{Paywall Polices}
Next, We measure the distribution of policies enforced by paywalls.  We find that paywalls vary widely by type, enforcement mechanism, and how much, if any, content visitors can view before needing to pay.  For these measurements, we randomly sample ~\ManuallyReviewedPaywallSites paywall-using websites from our dataset for manual evaluation.

\begin{figure*}[t]
    \centering 
    \begin{minipage}{0.32\linewidth}   
        \centering
        \includegraphics[width=1.05\linewidth]{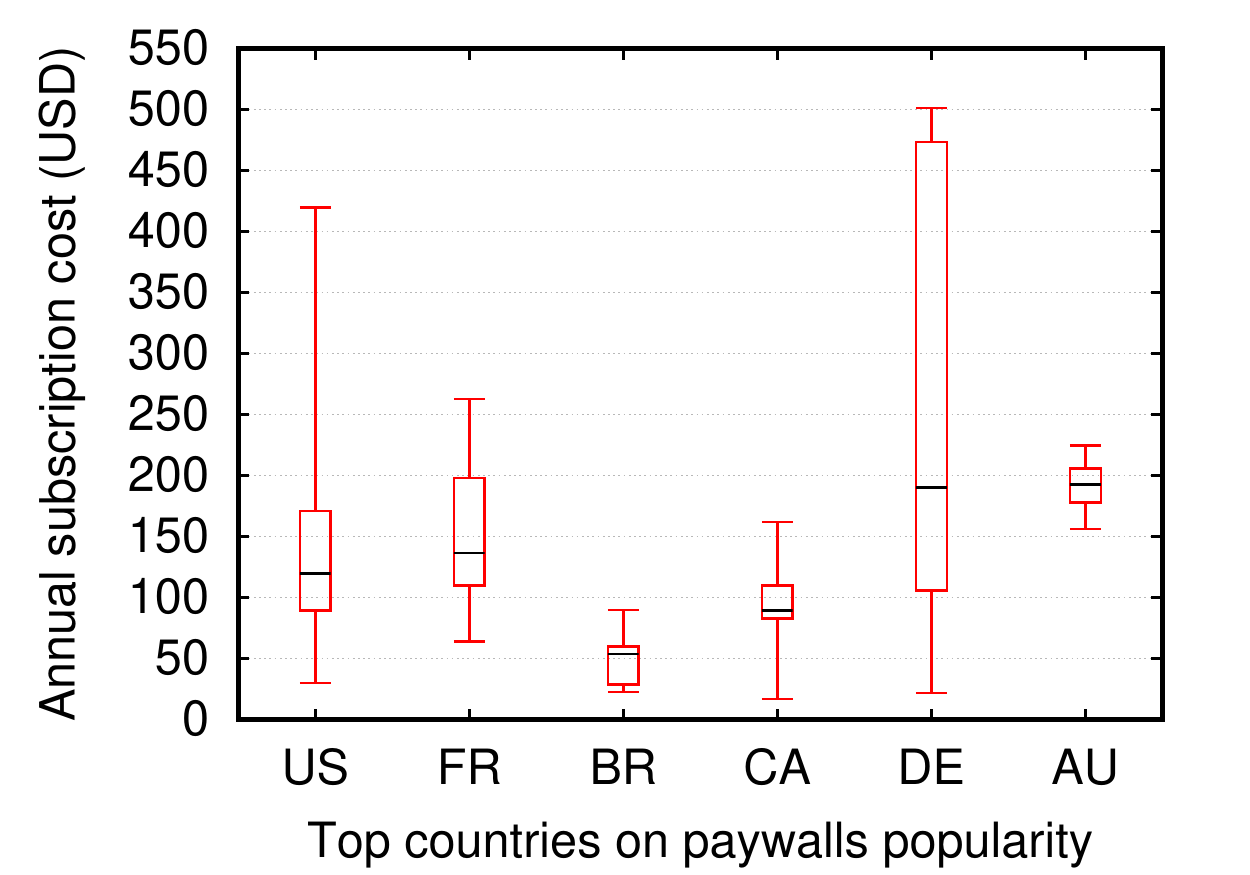}
        \caption{Min, 15th percentile, median, 85th percentile, and max annual subscription costs for paywalls, by country.}
        \label{fig:cost_per_country}
        \end{minipage}
    \hfill
        \begin{minipage}{0.33\linewidth} 
        \centering\vspace{0.2cm}
        \includegraphics[width=1.05\linewidth]{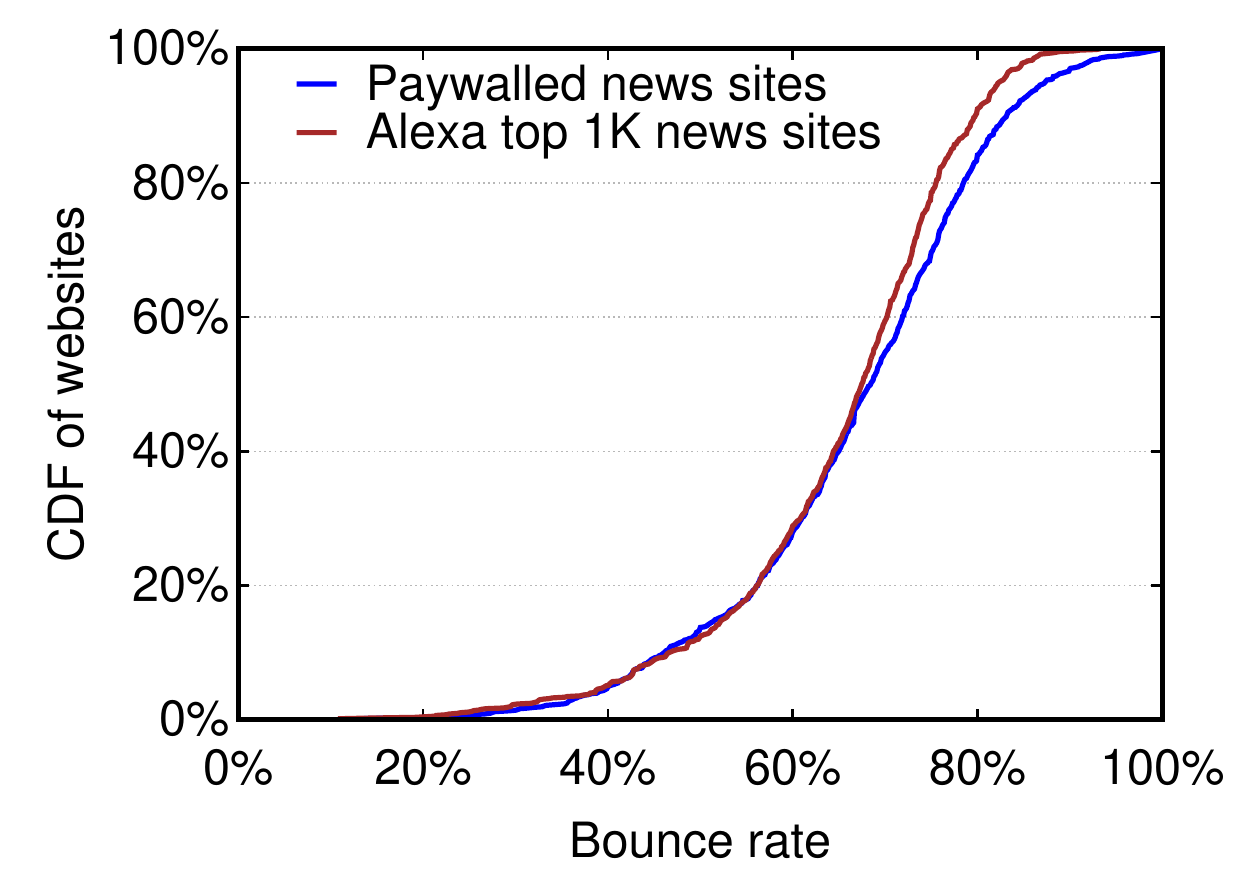}\vspace{-0.1cm}
        \caption{Distribution of the bounce rate per website. The median paywalled site has slightly higher bounce rate (\bounceRatePaywalled) contrary to the non-paywalled (\bounceRateNonPaywalled).}
        \label{fig:alexa_bounce_cdf}\vspace{-0.2cm}
    \end{minipage}
    \hfill
    \begin{minipage}{0.33\linewidth} 
        \centering\vspace{0.4cm}
        \includegraphics[width=1.05\linewidth]{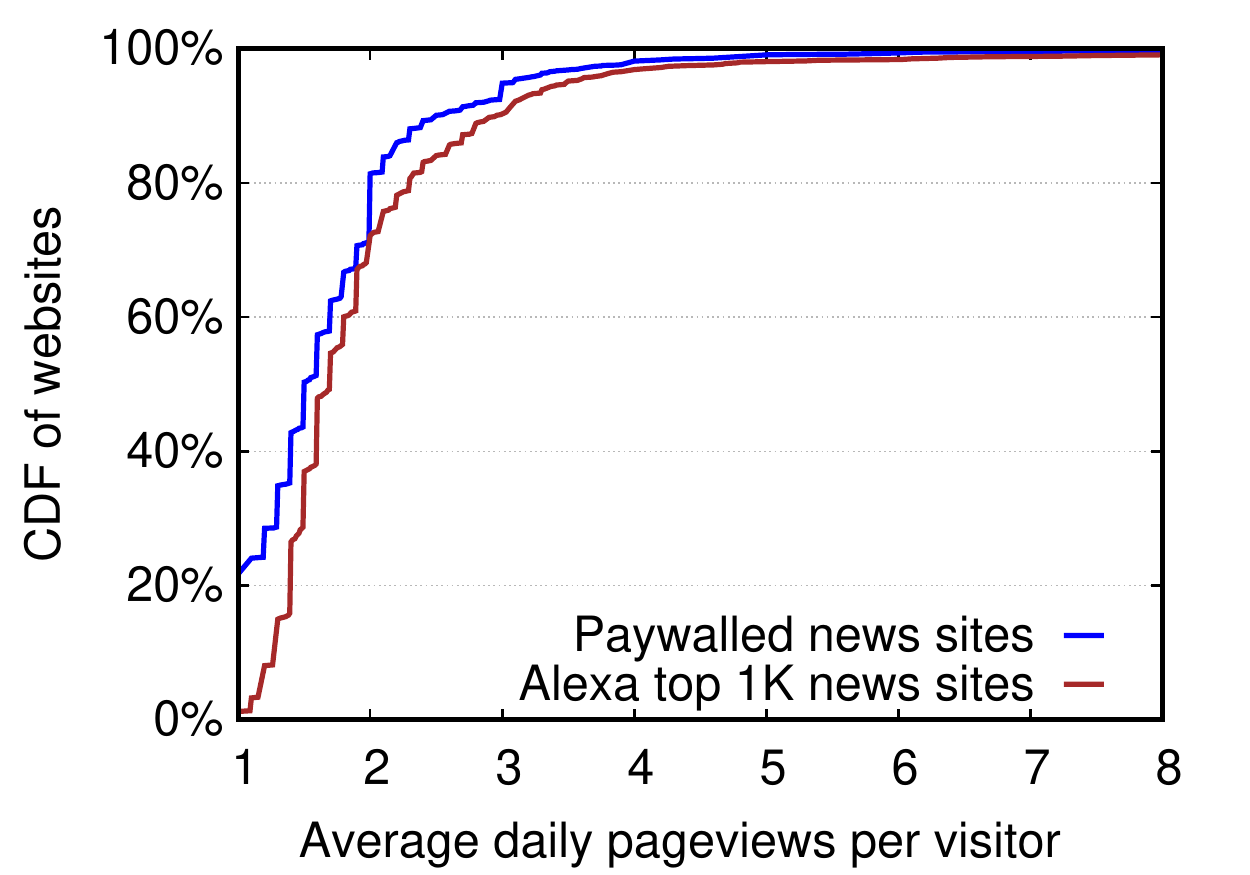}
        \caption{Distribution of the daily page views per visitor per news site. In median values, an average visitor browses on a daily basis \pageviewDiff less pages on a paywalled news site.}\vspace{-0.3cm}
        \label{fig:pageviews}
     \end{minipage}
\end{figure*}

\subsubsection{The different types of Paywalls} 
First, we observe that \softPaywalls of paywalls are ``soft'' (\ie allow some free content access), \hardPaywalls are ``hard'' (\ie allow no free access), with the remaining~\freemiumPaywalls paywalled sites using a ``hybrid'' strategy (\ie some content is free, some requires payment, based on the author/time of publication/topic, \etc). Some ``hybrid'' publishers use machine learning or other dynamic approaches to determine per-user whether an article should be locked or not~\cite{mlpaywall,vidoraML,algorithmicPaywall}.

\subsubsection{Enforcement Mechanism}
 We also measure the distribution of paywall enforcement techniques. Despite the heterogeneity of the paywall implementations, we see only three approaches used to enforce paywalls: (i) truncating article text, (ii) obfuscating the article with popups, or (iii) redirecting users to a subscription page. We measure the popularity of each of the above approaches in our manually evaluated set; Figure~\ref{fig:enforcingStrategies} presents the results. The largest percentage (\obfuscatedArticles) of the websites obfuscate or truncate~(\truncatedArticles) the article the user has not yet access to. Only a few~(\redirectedArticles) redirect the user to a login/subscribe page.

\subsubsection{Allowed Free Content}
 We also measure the distribution of how much content users can view before triggering a (hard or soft) paywall. For the ~15.7\% of sites that use a hard-paywall strategy, visitors cannot view \emph{any} articles for free.  For soft paywalls, this number varies by publisher. Figure~\ref{fig:freeArticles} plots the distribution of the free articles users could consume before hitting a paywall in the websites we tested. Overall, the median paywalled website allows \freeArticlesMedianAll articles. All hard paywalled websites do not allow \emph{any} access to articles, when the median soft-paywalled website allows \freeArticlesMedianSoft articles to be read for free. A significant number of soft paywalls~(30\%) that allow~2 or fewer articles to be read before triggering enforcement.
 
\subsubsection{Paywalls Cost}
\label{sec:cost}
Next, we measure the distribution of paywall subscription costs. We find that most paywall subscriptions are monthly, that the median annual cost for paywall access is 108 USD, and that subscription costs seem to be highest in Germany.  All of these measurements were conducted through a manual evaluation of \costSubscription paywalled sites.  We sampled 20 sites from each of the top~6 paywall using countries.  For 12 sites, we were not able to access the site or determine the subscription costs

We first measure the distribution of subscription options for users. ~\monthlySubs of paywall sites provide a monthly subscription option and ~\annualSubs of sites provide an annual one. Hence,~\monthlyOnlySubs of the paywalled sites provide \emph{only} a monthly subscription option and ~\annualOnlySubs \emph{only} an annual one.
Next, we measure the distribution of purchasing an annual subscription to a site's content.  The median observed annual subscription cost is 108 USD. ~22\% of sites charge less than 60 USD, ~21\% of sites charge more than~180 USD.  Figure~\ref{fig:cost_subscr_cdf} presents the full distribution of annual subscription costs.  We note that the subscription rates we observe are lower than those estimated by previous work (around~189 USD on average)~\cite{paywallNiemanlab}, possibly reflecting a general decrease in costs. We measure the distribution of annual costs by manually noting the annual subscription cost in the local currency.  For sites that do not offer an annual subscription, we sum the cost of twelve monthly subscriptions.  We then convert all costs to USD for comparison purposes.

Finally, we measure how subscription costs differ by country. As depicted in Figure~\ref{fig:cost_per_country}, we plot the min, the~15th percentile the median, the~85th percentile and the max of the annual subscription cost across the different countries. The median prices of subscriptions in Australia and Germany are highest (193 and 190 USD, respectively).  Subscription costs vary widely by site, too.  In Germany and the United States, for example, the most expensive paywalls cost ~$2.63\times$ and~$3.51\times$ more than the median rate, respectively.
\subsection{How Paywalls Affect Site Use}
\label{sec:alexa}
Paywalls affect how users interact with the site.  Recent studies~\cite{kim2019newspapers}, by monitoring the pageviews of 36 news sites before and after paywall deployment, report that they lose nearly~30\% of their daily traffic and a loss of pageviews, ranging from a~10\% to~55\%. 
In this section, we measure differences between how sites interact with paywalled and non-paywalled sites.  We find that users view less pages on paywalled sites, stay for shorter periods of time and link to pages less.  Interestingly, we did not see a significantly difference to the bounce rate between paywalled and non-paywalled sites~\footnote{We do not address the issue of causation; its possible, for example, that the types of site likely to use paywalls have lower \emph{dwell times} already, so that the use of a paywall is a more a result of lower dwell time than the cause.  We leave disentangling cause and effect for future work.}.

\begin{figure*}[t]
    \centering
    \begin{minipage}{0.32\linewidth}      
        \centering
    \includegraphics[width=1.05\linewidth]{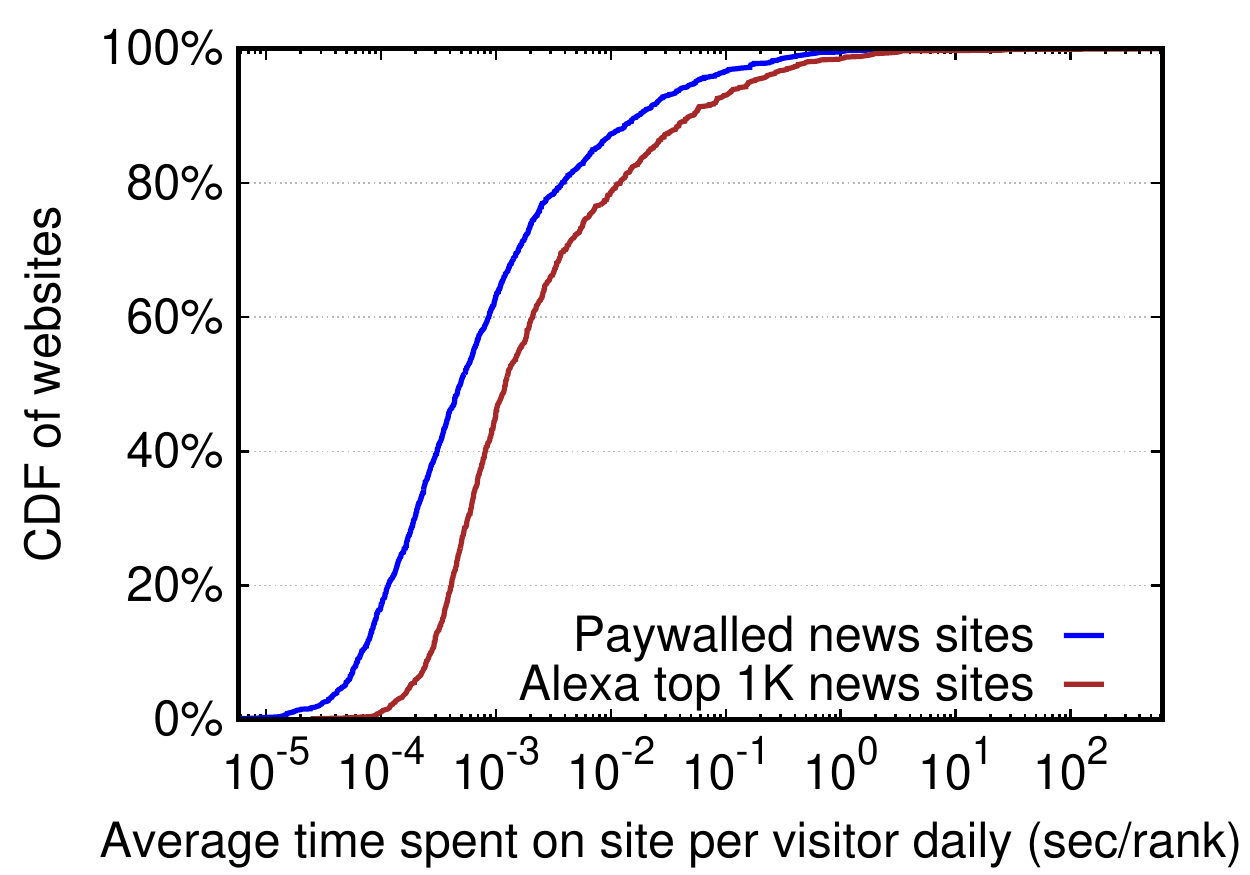}
    \caption{Distribution of the average time a visitor spends daily per news site. In median values visitors tend to spend daily \spentTimeDiff more time per site rank on non-paywalled websites.}
    \label{fig:dailyTimeOnSite}
    \end{minipage}
    \hfill
    \begin{minipage}{0.32\linewidth}
       \centering\vspace{-0.3cm}
        \includegraphics[width=1.05\linewidth]{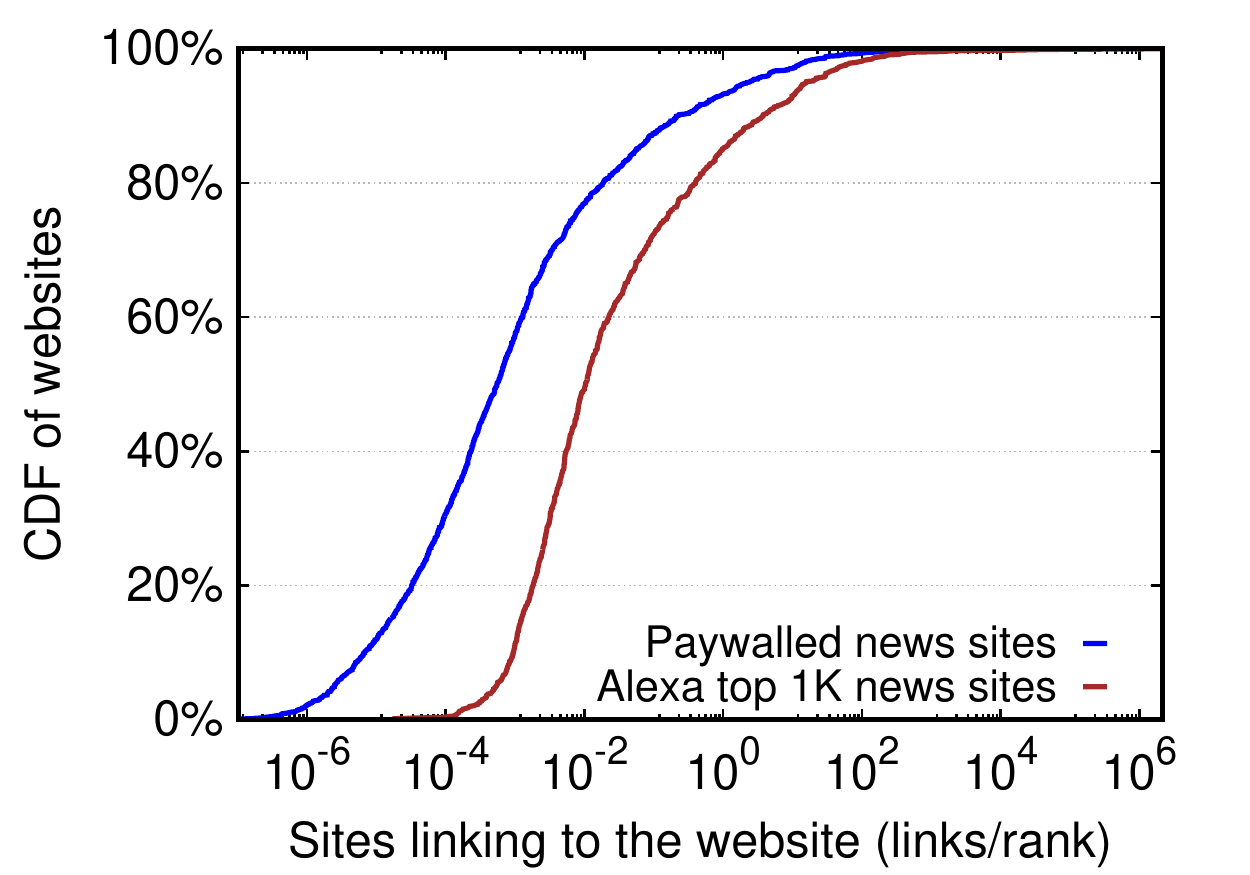}
        \caption{Distribution of the incoming site links per news site. Paywalled sites get significantly (\backLinkDiff) less site links per rank in median values.}
        \label{fig:siteLinks}
        \end{minipage}
    \hfill
    \begin{minipage}{0.32\linewidth}
        \centering
        \vspace{-.3cm}
        \includegraphics[width=1.05\linewidth]{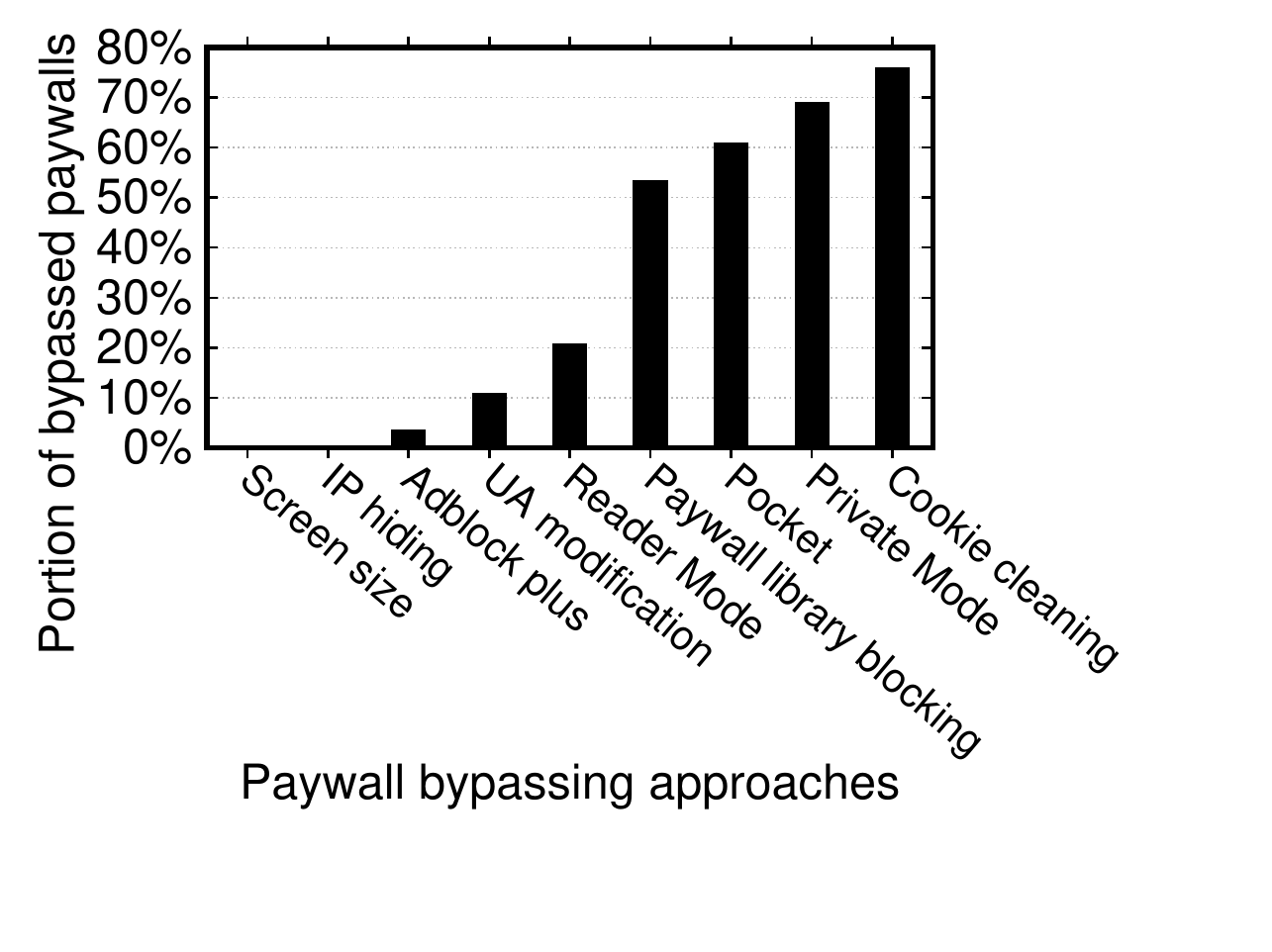}
        \vspace{-.6cm}
        \caption{Success rate of the different paywall bypassing approaches. Clearing the cookie jar alone can bypass~75\% of the paywalls.}
        \label{fig:paywallsBypassFeatures}
    \end{minipage}
\end{figure*}

\subsubsection{Bounce Rate}
We find that paywalled new sites have slightly higher bounce rates~\footnote{The percentage of visitors who visit a site and then leave, rather than continuing to view other pages within the same site.} than non-paywalled news sites.  The distributions of bounce rates is depicted in Figure~\ref{fig:alexa_bounce_cdf}.
The median paywalled news site has slightly higher bounce rate (\bounceRatePaywalled) contrary to the median non-paywalled (\bounceRateNonPaywalled). However, we see that for ~30\% of the websites in the two categories the difference is~2-7\% higher.  To compare bounce rates, we used the Alexa Top Sites data, and compared the bounce rates for the paywalled news sites in our data set with the Alexa top ~1K news sites.

\subsubsection{Daily Page Views}
Next, we measure the number of pages the average visitor performs daily on the websites and we compare how this changes for the paywalled and non-paywalled news sites. In Figure~\ref{fig:pageviews}, we plot the cumulative distribution of these page views per website in our dataset. Users visit on average \pageviewDiff less pages on paywalled news sites than non-paywalled new sites.

\subsubsection{Average Time spent on Site}
Figure~\ref{fig:dailyTimeOnSite} compares the distribution of the median time users spend on paywalled and non-paywalled websites, normalized by popularity (based on its Alexa rank). We find that visitors spend daily \spentTimeDiff more time per on non-paywalled news sites.

\subsubsection{Content Popularity and Link Rate}
Finally, we measure the impact of paywalls on how often sites link to the paywalled sites. Since site linking may be affected by the popularity of the website, in Figure~\ref{fig:siteLinks}, we plot the cumulative distribution of the number of site links (or backlinks) per news site normalized by its Alexa rank. We observed paywalled sites being linked to significantly less (\backLinkDiff) often than non-paywalled sites.
\subsection{Paywalls and Privacy}
\label{sec:privacy}
Most behavioral advertising systems require users to pay for content with their privacy; users are tracked in behavioral advertising systems, and can view ``free'' content.
Paywalls have the possibility of changing this system.  Since users are directly paying for content, one might hope users would no longer face the privacy harms associated with behavioral advertising systems.  Unfortunately, we see that this is not the case.  People \textit{do not} generally receive a tracker free version of site content when paying for subscriptions.  Instead, paywall systems seem to serve as an \textit{additional} monetization mechanism on top of existing, privacy harming, ad systems.

\begin{figure}[b]
\centering
\vspace{-0.2cm}
    {\small 
    \begin{tabular}{l|rr|rr}
        \toprule
         & \multicolumn{2}{c}{\bf  Vanilla User} & \multicolumn{2}{c}{\bf Premium User}  \\
         {\bf News site} & {\bf Ads} & {\bf Tracking} & {\bf Ads} & {\bf Tracking}\\ \midrule
            heraldsun.com.au & 171 & 13 & 169 & 9\\
            miamiherald.com & 123 & 12 & 112 & 11\\
            wsj.com & 63 & 4 & 61 & 4\\
            kansascity.com & 61 & 9 & 56 & 6 \\ 
            ft.com & 20 & 0 & 11 & 0\\
            salon.com & 138 & 5 & 0 & 1 \\
            japantimes.co.jp & 109 & 12 & 98 & 8 \\
            leparisien.fr & 125 & 10 & 81 & 4 \\
            independent.co.uk & 11 & 6 & 10 & 6 \\
            spectator.co.uk & 18 & 2 & 14 & 2 \\
            \bottomrule
    \end{tabular}\vspace{-0.2cm}
    \caption{Requests captured for vanilla and premium user. User continues receiving the same amount of trackers and ads in the content she receives even if she has paid for it.}
    \label{tbl:withAndWithout}
    }\vspace{-0.2cm}
\end{figure}

We measure whether paying for paywall access improves user privacy (\ie removes the need for sites to try and monetize through tracking) by purchasing subscriptions to \NumSubscriptions randomly selected paywalled news sites. Our goal is to examine the types of network requests issued before and after paying for the subscription. 
We create two scenarios, the vanilla (non-subscribed) user, and the premium (subscribed) user. For each selected site, we create an account and purchased a subscription before the starting the measurement.  We also select 5 child pages on each site for evaluation.
Then, we enable the popular Disconnect plugin~\cite{disconnect} in monitoring and no-blocking mode, and browse each selected child page on each site under each of the two personas, in the same order, and observe the issued network requests. Figure~\ref{tbl:withAndWithout} presents the average number of ad- and tracking- related requests encountered in each scenario. In the vast majority of cases, there is no significant difference in terms of ad- or tracking- related web requests.
\section{Paywall Circumvention}
\label{sec:circumvention}

Paywalls must be robust to circumvention if they are going to be a successful monetization scheme for websites.  If paywalls can be easily avoided, then content producers will wind up in the same situation they are in with ads and ad-blockers; declining revenues as circumvention tools become more popular.  We find that \textit{all observed paywalls are trivial to circumvent}.

We evaluate how robust paywalls are to circumvention in two steps: (i) we categorize the approaches of several popular paywall circumvention strategies, and (ii) we test each strategy on \circumventSet paywalled news sites, we randomly select from our dataset. This  subset comprises~28 soft and~4 hard paywalls on popular websites like Wired, Bloomberg, Spectator, Washington Post, Irish Times, Medium, Build, Japan Times, Statesman, and Le Parisien.

\subsection{Evasion Evaluated}
We test the robustness of each paywall system by using Chrome version~71.  For each evaluated site, (a) we browse different pages till we trigger the paywall, and then (b) we test a variety of bypassing approaches to circumvent the paywall and get access to the ``protected'' article. 
Figure~\ref{fig:paywallsBypassFeatures} lists the evaluated paywall-circumvention strategies, which includes pre-packaged tools, fingerprint evasion techniques, and third-party services.  Specifically, we consider:
\begin{enumerate}\itemsep=-1pt
\item changing the screen size dimensions
\item hiding the user's IP address
\item changing the user agent string
\item using an ad blocker extension
\item enabling ``Reader Mode''
\item using the Pocket web service\footnote{or similar ``reader'' services like ``JustRead''~\cite{justRead} and ``Outline''~\cite{outline}}
\item enabling Incognito/Private Mode
\item emptying the cookie jar
\item blocking HTTP requests for popular paywall libraries
\end{enumerate}
Overall, we are able to bypass \emph{all of the soft paywalls and none of the hard paywalls}. Hard paywalls perform their enforcement server-side, when the soft paywalls perform their policy enforcement client-side, and thus their access control is circumventable.

\begin{figure}[tb]
    \centering
    \vspace{-0.3cm}
    \includegraphics[width=0.65\linewidth]{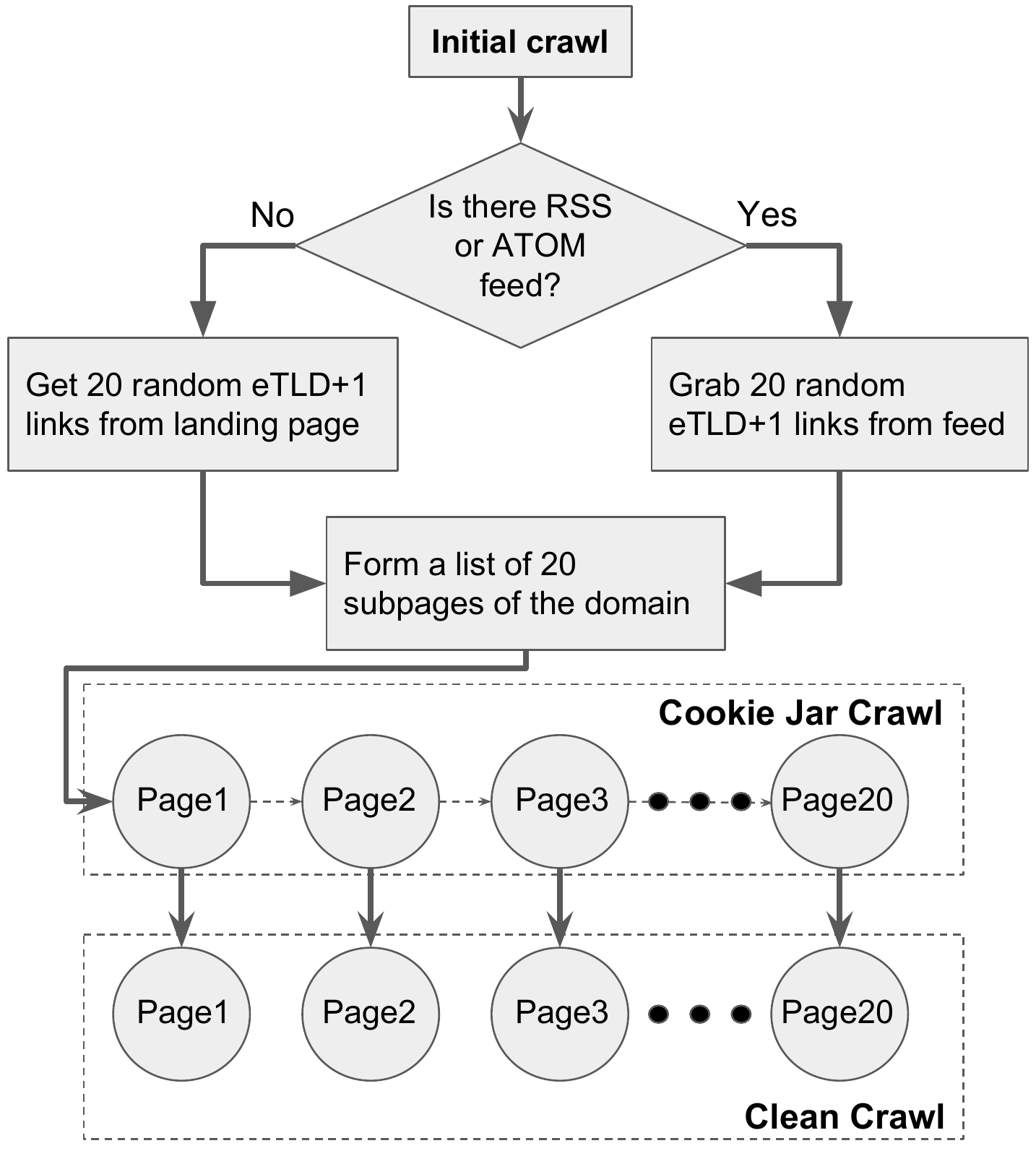}
    \vspace{-0.2cm}
    \caption{Data collection steps of our paywall detector's crawling component. }
    \vspace{-0.5cm}
    \label{fig:visitedSites}
\end{figure}

\subsection{Evasion Approaches Analyzed}
Many of the evaluated evasion approaches are rarely successful. For example, changing the screen size or the IP address of the user rarely circumvents a soft paywall~(4\% effectiveness). A moderate number of soft paywalls~(12\%) is flummoxed by modifying the browser's user agent string.
The majority~(75\%) of soft paywalls is bypassable by resetting the cookie jar (in some cases erasing the first-party cookie only is insufficient, since it is automatically re-spawned by fingerprinting JavaScript code, as seen in Section~\ref{sec:casestudy}). As a result, switching into browsers' ``private browsing'' modes is also sufficient to bypass most paywalls.  Some paywalled sites refuse to render content in ``reader modes'' or ``private browsing'' modes, either first party (\eg reader modes shipped with Safari and Firefox) or third-party (\eg services like Pocket~\cite{pocket}). Such detection schemes are uncommon though; switching into reader-mode, for example, circumvents paywall enforcement in~60\% of the cases.
Ad-blocking extensions, in their default configurations, have little-to-no effect on paywalls. However, by using the list of known paywall libraries from Section~\ref{sec:results} and by blocking requests to these domains we are able to bypass~48\% of the paywalls without breaking the website's main functionality.

Third-parties like Google Search, Twitter, Reddit and Facebook, can also be used to gain access to some paywalled articles. Some paywalls give visitors from these large third-party systems unfettered access to their content, in pay-for-promotion initiatives. By spoofing the referrer field of the HTTP GET requests, some paywalls are vulnerable to exploiting a controversial policy~\cite{firstclick} where publishers (for promotion purposes) allow access to articles when the visitor comes from one of these platforms (by clicking on a tweet, a post, a Google search result \etc)~\cite{referrerBypass}. These mechanisms can provide access to hard paywalled articles. As a result, some publishers (\eg Wall Street Journal) have stopped allowing such special access through their paywalls~\cite{wsJGoogleRef}.
\section{Paywall Detection}
\label{sec:detection}

\begin{figure}[t]
    \centering
    {\small
    \vspace{-0.2cm}
    \begin{tabular}{lr}
        \toprule
        {\bf Metric} & {\bf Value}  \\ \midrule
        Precision & \ClassifierEvalPrecision{} \\
        Recall  & \ClassifierEvalRecall{} \\
        F-Measure & \ClassifierEvalFMeasure \\
        AUROC & \ClassifierEvalROCAUC \\
        \bottomrule
    \end{tabular} \vspace{-0.2cm}
    }
    \caption{Weighted average of the performance of our RF classifier, after k=5 cross-fold validation.}
    \label{table:model-performance}
    \vspace{-0.5cm}
\end{figure}

This section presents the design and evaluation of a ML-based detection system
whose goal is to determine whether a site uses a paywall. Our paywall detector
consists of two components: (i) a crawling component that visits a subset of
pages on a site, records information about each page's execution, and extracts
some ML features; and (ii) a classifier, that uses the extracted features to
predict if the site uses a paywall.

We present this classifier as a partial solution for the problem of measuring
changes in the adoption and behavior of paywalls over time.  We propose this ML
approach as a \emph{complement} to the crowd-sourced approach described in
Section~\ref{sec:dataset}. The classifier can be used to automatically gauge
paywall prevalence. This automated approach can help identify and quantify
paywalls that have not been identified by crowd-sourced
lists, such as paywalls deployed by unpopular or region-specific sites.

Before describing the classifier in detail, we note two things. First,
the classifier is designed to help detect broad, web-scale trends in
paywall use and behavior, not to detect at real time paywall use on any single specific site.
Second, an important finding of this classifier is that there is far greater
diversity in paywall behavior and implementation logic than we expected at the
start of the effort.  We expect this to be a useful starting point for future 
studies.

\subsection{Crawling Methodology}
\label{sec:methodology-crawling}
The data collection step of our paywall detector, depicted in
Figure~\ref{fig:visitedSites}, begins with three crawls of the target website:
(i) the \emph{initial crawl} that collects a list of child pages on the website,
(ii) the \emph{cookie jar crawl}, where each child page is crawled
sequentially in the same browsing session and (iii) the \emph{clean
crawl}, where each child page is crawled with a fresh browsing session (\ie a
``clean'' cookie jar).

This strategy replicates viewing patterns that might cause a paywall to be
triggered, and then attempts to detect the paywall's presence by looking for
page content that was visible on previous visits, but is no longer visible.  For
each page crawled, the crawler records the final state of the DOM, which DOM
elements are visible, which page elements are positioned to obscure others (\eg
modal dialogs), and other page execution data only available at runtime.

\subsection{Feature Extraction}
We selected features that target both immediately triggering paywalls and
paywalls that trigger after viewing multiple pages. These features aim to
capture an intuition about how paywalls behave, and can fall into three rough
categories: textual features, structural features, and visual features.

\point{Text features} These features consider the text of the page, targeting
text and idioms associated with paywalls.  The crawler looks for the phrases
``subscribe'', ``sign up'' and ``remaining'' (translated into 87 languages) in
(i) the \RM subset of the page, (ii) any overlay or popup elements (\eg
elements that have, or are children of elements that have, z-index values
greater than zero), and (iii) elsewhere in the page. These three checks
are performed both in \CJq and the \CCq recordings of each page. 

Several text features use a \RM version of page, the subsection of the document
identified as the page's ``main content'', or the content stripped of page
``boilerplate'' elements (\eg advertisements, navigation elements, decorative
images).  While there are many different \RM identification
strategies~\cite{ghasemisharif2018speedreader}, in this work we use Mozilla's
\emph{Readability.js}~\cite{readability} implementation, because of its
popularity and ease of use.  We expect using other \RM strategies would work
roughly as well.

\point{Structural features} 
These features target page structure (\ie HTML), independent
of specific page text or presentation. Structural features include whether the
website has a RSS or ATOM feed, changes in the number of text nodes present in
the page between its \CJq and \CCq versions, how many
measured pages contain a \RM subset, and the average and maximum
difference in the amount of text in the document in \RM, between \CJq and \CCq
measurements.

\point{Visual features} 
These features focus on visual aspects of measured pages, and how those visual
aspects change between the \CJq and \CCq measurements for each child page. The
detector measures how many text nodes are obscured and the average and maximum
change in obscured text nodes between the two measurements for each page.
Additional display features are the number, and change in, text nodes in the
browser viewport, and number of text nodes (regardless of text content)
appearing in overlay (\ie $z$-index great than zero) page elements. These
features identify paywalls that prevent users from reading page content
through popups or similar methods.

\subsection{Classifier Accuracy}
\label{sec:classification}
Our paywall detector uses a \emph{random forest} classifier, specifically the
\textit{RandomForestClassifier} implementation provided by the popular
SciKit-Learn~\cite{sklearn} python library. Classification parameters were
selected through~5-fold evaluation using the entirety of the aforementioned
extracted features. As a ground truth, we use a subset of the paywalled sites the oracle identified (Section~\ref{sec:dataset}). To assess the accuracy of the classifier we use a different subset of the oracle's data and a set of non-paywalled websites we manually generate. The paywall detector achieves an 
average precision of~\ClassifierEvalPrecision, recall of~\ClassifierEvalRecall and an area 
under the receiver operating characteristics~(AUROC) of~\ClassifierEvalROCAUC. These
results are encouraging and suggest that our approach can be used to gauge
paywall prevalence on the web. They also indicate that paywalls
vary in behavior more than we anticipated, and that more complex features
may be needed to further improve accuracy.

\section{Related Work} 
\label{RelatedWork}

In~\cite{cornia2017pay}, authors perform an empirical study of the pay models (freemium and paywall models) in European news. In particular, they manually analyzed a small dataset of~171 of the most important news organizations in France, Poland, Germany, Italy, Finland, and UK. Their results show that~66\% percent of the  newspapers operate a pay model and that the average price for a monthly subscription is~13.64 Euros when prices in general range from~2.10 to~54.27 Euros/month. In our measurements, 3 years after, the average monthly subscription cost~10.93 Euros, when specifically in Germany it is 20.48 (was 19.75) Euros and in France it is 12.54 (was 13.97) Euros.

In~\cite{myllylahti2017content}, authors explore the content that news publishers consider worthy of placing behind a paywall. They analyze~614 articles from the leading Australasian financial newspapers (\ie the Australian Financial Review (AFR) and the National Business Review (NBR)). Results show that publishers consider hard (or fast-paced) news and opinion pieces as the most valuable news commodity. In addition, as presented, AFR locked~86\% of its content compared to NBR's~41\%.  

In~\cite{carson2015behind}, authors analyze selected paywalled news sites in US, UK and Australia to compare the type, pricing and audience uptake. Results show that paywalls are part of newspapers' toolkit for bringing in new revenue but there is no evidence to suggest they can be a standalone solution. However, in this political economic environment for mastheads, digital advertising revenues alone are also insufficient to meet the cost of providing quality journalism.
\section{Conclusion} 
\label{sec:conclusion} 

Despite the seemingly important implications, paywalls impose on the free web, as an internet phenomena, they have been understudied. This paper aims to address this blind spot by conducting the first large scale study of paywalls on the web. Our results show that paywall use increases over time ($2\times$ more paywalls every 6 months), its adoption differs by country (\eg \paywallAdoptionUS in US, \paywallAdoptionAU in Australia), and besides the privacy implications, paywalls fail to reliably protect publishers content. Finally, we present the design of a novel, automated system for detecting whether a site uses a paywall. We hope this work can be a significant first step in understanding the phenomena of paywalls.

\bibliographystyle{ACM-Reference-Format}
\balance
\bibliography{paper}


\begin{thebibliography}{64}


\ifx \showCODEN    \undefined \def \showCODEN     #1{\unskip}     \fi
\ifx \showDOI      \undefined \def \showDOI       #1{#1}\fi
\ifx \showISBNx    \undefined \def \showISBNx     #1{\unskip}     \fi
\ifx \showISBNxiii \undefined \def \showISBNxiii  #1{\unskip}     \fi
\ifx \showISSN     \undefined \def \showISSN      #1{\unskip}     \fi
\ifx \showLCCN     \undefined \def \showLCCN      #1{\unskip}     \fi
\ifx \shownote     \undefined \def \shownote      #1{#1}          \fi
\ifx \showarticletitle \undefined \def \showarticletitle #1{#1}   \fi
\ifx \showURL      \undefined \def \showURL       {\relax}        \fi
\providecommand\bibfield[2]{#2}
\providecommand\bibinfo[2]{#2}
\providecommand\natexlab[1]{#1}
\providecommand\showeprint[2][]{arXiv:#2}

\bibitem[\protect\citeauthoryear{{Adam}}{{Adam}}{2018}]%
        {bypassExt4}
\bibfield{author}{\bibinfo{person}{{Adam}}.} \bibinfo{year}{2018}\natexlab{}.
\newblock \bibinfo{title}{Bypass Paywalls for Chrome}.
\newblock
  \bibinfo{howpublished}{\url{https://github.com/iamadamdev/bypass-paywalls-chrome}}.
\newblock


\bibitem[\protect\citeauthoryear{Appleby}{Appleby}{2016}]%
        {murmur3}
\bibfield{author}{\bibinfo{person}{Andy Appleby}.}
  \bibinfo{year}{2016}\natexlab{}.
\newblock \bibinfo{title}{MurmurHash3}.
\newblock
  \bibinfo{howpublished}{\url{https://github.com/aappleby/smhasher/wiki/MurmurHash3}}.
\newblock


\bibitem[\protect\citeauthoryear{Bilton}{Bilton}{2018}]%
        {wiredPaywall}
\bibfield{author}{\bibinfo{person}{Ricardo Bilton}.}
  \bibinfo{year}{2018}\natexlab{}.
\newblock \bibinfo{title}{Learning from the New Yorker, Wired's new paywall
  aims to build a more ''stable financial future''}.
\newblock
  \bibinfo{howpublished}{\url{http://www.niemanlab.org/2018/02/learning-from-the-new-yorker-wireds-new-paywall-aims-to-build-a-more-stable-financial-future/}}.
\newblock


\bibitem[\protect\citeauthoryear{Bonnie}{Bonnie}{2017}]%
        {firstclick}
\bibfield{author}{\bibinfo{person}{Christian Bonnie}.}
  \bibinfo{year}{2017}\natexlab{}.
\newblock \bibinfo{title}{Google's latest move means you actually have to pay
  for news}.
\newblock
  \bibinfo{howpublished}{\url{https://www.wired.co.uk/article/google-ditches-first-click-free-policy}}.
\newblock


\bibitem[\protect\citeauthoryear{Brian~Kennish}{Brian~Kennish}{2019}]%
        {disconnect}
\bibfield{author}{\bibinfo{person}{Casey~Oppenheim Brian~Kennish}.}
  \bibinfo{year}{2019}\natexlab{}.
\newblock \bibinfo{title}{Disconnect Browser plugin}.
\newblock \bibinfo{howpublished}{\url{https://disconnect.me}}.
\newblock


\bibitem[\protect\citeauthoryear{Brinkmann}{Brinkmann}{2016}]%
        {referrerBypass}
\bibfield{author}{\bibinfo{person}{Martin Brinkmann}.}
  \bibinfo{year}{2016}\natexlab{}.
\newblock \bibinfo{title}{Read articles behind paywalls by masquerading as
  Googlebot}.
\newblock
  \bibinfo{howpublished}{\url{https://www.ghacks.net/2016/02/26/read-articles-behind-paywalls-by-masquerading-as-googlebot/}}.
\newblock


\bibitem[\protect\citeauthoryear{Brown}{Brown}{2019}]%
        {fanboyPaywalls}
\bibfield{author}{\bibinfo{person}{Ryan Brown}.}
  \bibinfo{year}{2019}\natexlab{}.
\newblock \bibinfo{title}{Fanboy's Enhanced Tracking List}.
\newblock
  \bibinfo{howpublished}{\url{https://github.com/ryanbr/fanboy-adblock/blob/master/enhancedstats-addon.txt}}.
\newblock


\bibitem[\protect\citeauthoryear{Burgi}{Burgi}{2016}]%
        {adfraud}
\bibfield{author}{\bibinfo{person}{Michael Burgi}.}
  \bibinfo{year}{2016}\natexlab{}.
\newblock \bibinfo{title}{What's Being Done to Rein In \$7 Billion in Ad
  Fraud}.
\newblock
  \bibinfo{howpublished}{\url{https://www.adweek.com/brand-marketing/whats-being-done-rein-7-billion-ad-fraud-169743/}}.
\newblock


\bibitem[\protect\citeauthoryear{Burrell}{Burrell}{2018}]%
        {ftPaywall}
\bibfield{author}{\bibinfo{person}{Ian Burrell}.}
  \bibinfo{year}{2018}\natexlab{}.
\newblock \bibinfo{title}{The FT will next year hit 1m subscribers, 17 years
  after putting up its paywall}.
\newblock
  \bibinfo{howpublished}{\url{https://www.thedrum.com/opinion/2018/08/30/the-ft-will-next-year-hit-1m-subscribers-17-years-after-putting-up-its-paywall}}.
\newblock


\bibitem[\protect\citeauthoryear{Carson}{Carson}{2015}]%
        {carson2015behind}
\bibfield{author}{\bibinfo{person}{Andrea Carson}.}
  \bibinfo{year}{2015}\natexlab{}.
\newblock \showarticletitle{Behind the newspaper paywall--lessons in charging
  for online content: a comparative analysis of why Australian newspapers are
  stuck in the purgatorial space between digital and print}.
\newblock \bibinfo{journal}{\emph{Media, Culture \& Society}}
  \bibinfo{volume}{37}, \bibinfo{number}{7} (\bibinfo{year}{2015}),
  \bibinfo{pages}{1022--1041}.
\newblock


\bibitem[\protect\citeauthoryear{Conley}{Conley}{2018}]%
        {paywallsDemocracy2}
\bibfield{author}{\bibinfo{person}{Nicholas Conley}.}
  \bibinfo{year}{2018}\natexlab{}.
\newblock \bibinfo{title}{The Problem with Paywalls}.
\newblock
  \bibinfo{howpublished}{\url{https://nicholasconley.wordpress.com/2018/05/01/the-problem-with-paywalls/}}.
\newblock


\bibitem[\protect\citeauthoryear{Cornia, Sehl, Simon, and Nielsen}{Cornia
  et~al\mbox{.}}{2017}]%
        {cornia2017pay}
\bibfield{author}{\bibinfo{person}{Alessio Cornia}, \bibinfo{person}{Annika
  Sehl}, \bibinfo{person}{Felix Simon}, {and} \bibinfo{person}{Rasmus~Kleis
  Nielsen}.} \bibinfo{year}{2017}\natexlab{}.
\newblock \showarticletitle{Pay models in European news}.
\newblock \bibinfo{journal}{\emph{Reuters Institute for the Study of
  Journalism, University of Oxford}} (\bibinfo{year}{2017}).
\newblock


\bibitem[\protect\citeauthoryear{Daigniere}{Daigniere}{2017}]%
        {bypassExt1}
\bibfield{author}{\bibinfo{person}{Florent Daigniere}.}
  \bibinfo{year}{2017}\natexlab{}.
\newblock \bibinfo{title}{A browser extension that maximizes the chances of
  bypassing paywalls}.
\newblock
  \bibinfo{howpublished}{\url{https://github.com/nextgens/anti-paywall}}.
\newblock


\bibitem[\protect\citeauthoryear{Dangu}{Dangu}{2018}]%
        {malvertising}
\bibfield{author}{\bibinfo{person}{Jerome Dangu}.}
  \bibinfo{year}{2018}\natexlab{}.
\newblock \bibinfo{title}{Uncovering 2017’s Largest Malvertising Operation}.
\newblock
  \bibinfo{howpublished}{\url{https://blog.confiant.com/uncovering-2017s-largest-malvertising-operation-b84cd38d6b85}}.
\newblock


\bibitem[\protect\citeauthoryear{Davies}{Davies}{2019}]%
        {adfraud2}
\bibfield{author}{\bibinfo{person}{Jessica Davies}.}
  \bibinfo{year}{2019}\natexlab{}.
\newblock \bibinfo{title}{Ghost sites, domain spoofing, fake apps: A guide to
  knowing your ad fraud}.
\newblock
  \bibinfo{howpublished}{\url{https://digiday.com/media/ghost-sites-domain-spoofing-fake-apps-guide-knowing-ad-fraud/}}.
\newblock


\bibitem[\protect\citeauthoryear{Dunn}{Dunn}{2016}]%
        {wikipediaFunds1}
\bibfield{author}{\bibinfo{person}{Jeff Dunn}.}
  \bibinfo{year}{2016}\natexlab{}.
\newblock \bibinfo{title}{Wikipedia is asking for donations again — here’s
  how much cash it already has in the bank}.
\newblock
  \bibinfo{howpublished}{\url{https://www.businessinsider.com/wikipedia-donations-profit-money-chart-2016-11}}.
\newblock


\bibitem[\protect\citeauthoryear{Edmund~Lee}{Edmund~Lee}{2018}]%
        {nyTimesPaywall}
\bibfield{author}{\bibinfo{person}{Rani~Molla Edmund~Lee}.}
  \bibinfo{year}{2018}\natexlab{}.
\newblock \bibinfo{title}{The New York Times digital paywall business is
  growing as fast as Facebook and faster than Google}.
\newblock
  \bibinfo{howpublished}{\url{https://www.recode.net/2018/2/8/16991090/new-york-times-digital-paywall-business-growing-fast-facebook-google-newspaper-subscription}}.
\newblock


\bibitem[\protect\citeauthoryear{{eMarketer}}{{eMarketer}}{2017}]%
        {adDuopoly}
\bibfield{author}{\bibinfo{person}{{eMarketer}}.}
  \bibinfo{year}{2017}\natexlab{}.
\newblock \bibinfo{title}{Google and Facebook Tighten Grip on US Digital Ad
  Market}.
\newblock
  \bibinfo{howpublished}{\url{https://www.emarketer.com/Article/Google-Facebook-Tighten-Grip-on-US-Digital-Ad-Market/1016494}}.
\newblock


\bibitem[\protect\citeauthoryear{Firn}{Firn}{2018}]%
        {vidoraML}
\bibfield{author}{\bibinfo{person}{Michael Firn}.}
  \bibinfo{year}{2018}\natexlab{}.
\newblock \bibinfo{title}{Optimize Dynamic Paywall Conversions with Machine
  Learning}.
\newblock
  \bibinfo{howpublished}{\url{https://www.vidora.com/product-updates/dynamic-paywalls/}}.
\newblock


\bibitem[\protect\citeauthoryear{Ghasemisharif, Snyder, Aucinas, and
  Livshits}{Ghasemisharif et~al\mbox{.}}{2018}]%
        {ghasemisharif2018speedreader}
\bibfield{author}{\bibinfo{person}{Mohammad Ghasemisharif},
  \bibinfo{person}{Peter Snyder}, \bibinfo{person}{Andrius Aucinas}, {and}
  \bibinfo{person}{Benjamin Livshits}.} \bibinfo{year}{2018}\natexlab{}.
\newblock \showarticletitle{SpeedReader: Reader Mode Made Fast and Private}.
\newblock \bibinfo{journal}{\emph{arXiv preprint arXiv:1811.03661}}
  (\bibinfo{year}{2018}).
\newblock


\bibitem[\protect\citeauthoryear{Greenslade}{Greenslade}{2013}]%
        {hardVsSoft}
\bibfield{author}{\bibinfo{person}{Roy Greenslade}.}
  \bibinfo{year}{2013}\natexlab{}.
\newblock \bibinfo{title}{Soft paywalls retain more users than hard paywalls -
  by a big margin}.
\newblock
  \bibinfo{howpublished}{\url{https://www.theguardian.com/media/greenslade/2014/nov/07/paywalls-charging-for-content}}.
\newblock


\bibitem[\protect\citeauthoryear{{Internet Archive}}{{Internet
  Archive}}{2001}]%
        {wayback}
\bibfield{author}{\bibinfo{person}{{Internet Archive}}.}
  \bibinfo{year}{2001}\natexlab{}.
\newblock \bibinfo{title}{Wayback Machine}.
\newblock \bibinfo{howpublished}{\url{https://archive.org/web/}}.
\newblock


\bibitem[\protect\citeauthoryear{Kim, Song, and Kim}{Kim et~al\mbox{.}}{2019}]%
        {kim2019newspapers}
\bibfield{author}{\bibinfo{person}{Ho Kim}, \bibinfo{person}{Reo Song}, {and}
  \bibinfo{person}{Youngsoo Kim}.} \bibinfo{year}{2019}\natexlab{}.
\newblock \showarticletitle{Newspapers’ Content Policy and the Effect of
  Paywalls on Pageviews}.
\newblock \bibinfo{journal}{\emph{Journal of Interactive Marketing}}
  (\bibinfo{year}{2019}).
\newblock


\bibitem[\protect\citeauthoryear{{kufii}}{{kufii}}{2019}]%
        {bypassExt3}
\bibfield{author}{\bibinfo{person}{{kufii}}.} \bibinfo{year}{2019}\natexlab{}.
\newblock \bibinfo{title}{Newspaper Paywall Bypasser}.
\newblock
  \bibinfo{howpublished}{\url{https://greasyfork.org/en/scripts/18585-newspaper-paywall-bypasser/code}}.
\newblock


\bibitem[\protect\citeauthoryear{Lawler}{Lawler}{2018}]%
        {paywallsApple}
\bibfield{author}{\bibinfo{person}{Richard Lawler}.}
  \bibinfo{year}{2018}\natexlab{}.
\newblock \bibinfo{title}{Apple seeks major newspaper allies for its
  subscription bundle}.
\newblock
  \bibinfo{howpublished}{\url{https://www.engadget.com/2018/09/08/apple-seeks-major-newspaper-allies-for-its-subscription-bundle/}}.
\newblock


\bibitem[\protect\citeauthoryear{Leung, Ren, Choffnes, and Wilson}{Leung
  et~al\mbox{.}}{2016}]%
        {Leung:2016:YUA:2987443.2987456}
\bibfield{author}{\bibinfo{person}{Christophe Leung}, \bibinfo{person}{Jingjing
  Ren}, \bibinfo{person}{David Choffnes}, {and} \bibinfo{person}{Christo
  Wilson}.} \bibinfo{year}{2016}\natexlab{}.
\newblock \showarticletitle{Should You Use the App for That?: Comparing the
  Privacy Implications of App- and Web-based Online Services}. In
  \bibinfo{booktitle}{\emph{Proceedings of the 2016 Internet Measurement
  Conference}} \emph{(\bibinfo{series}{IMC '16})}. \bibinfo{publisher}{ACM},
  \bibinfo{address}{New York, NY, USA}, \bibinfo{pages}{365--372}.
\newblock
\showISBNx{978-1-4503-4526-2}
\urldef\tempurl%
\url{https://doi.org/10.1145/2987443.2987456}
\showDOI{\tempurl}


\bibitem[\protect\citeauthoryear{Liu, Nath, Govindan, and Liu}{Liu
  et~al\mbox{.}}{2014}]%
        {Liu:2014:DDC:2616448.2616455}
\bibfield{author}{\bibinfo{person}{Bin Liu}, \bibinfo{person}{Suman Nath},
  \bibinfo{person}{Ramesh Govindan}, {and} \bibinfo{person}{Jie Liu}.}
  \bibinfo{year}{2014}\natexlab{}.
\newblock \showarticletitle{DECAF: Detecting and Characterizing Ad Fraud in
  Mobile Apps}. In \bibinfo{booktitle}{\emph{Proceedings of the 11th USENIX
  Conference on Networked Systems Design and Implementation}}
  \emph{(\bibinfo{series}{NSDI'14})}. \bibinfo{publisher}{USENIX Association},
  \bibinfo{address}{Berkeley, CA, USA}, \bibinfo{pages}{57--70}.
\newblock
\showISBNx{978-1-931971-09-6}
\urldef\tempurl%
\url{http://dl.acm.org/citation.cfm?id=2616448.2616455}
\showURL{%
\tempurl}


\bibitem[\protect\citeauthoryear{Masnick}{Masnick}{2018}]%
        {techdirtAdLoss}
\bibfield{author}{\bibinfo{person}{Mike Masnick}.}
  \bibinfo{year}{2018}\natexlab{}.
\newblock \bibinfo{title}{The Media's Paywall Obsession Will End In Disaster
  For Most}.
\newblock
  \bibinfo{howpublished}{\url{https://www.techdirt.com/articles/20180506/11501539779/medias-paywall-obsession-will-end-disaster-most.shtml}}.
\newblock


\bibitem[\protect\citeauthoryear{McCarthy}{McCarthy}{2019}]%
        {drum2019}
\bibfield{author}{\bibinfo{person}{John McCarthy}.}
  \bibinfo{year}{2019}\natexlab{}.
\newblock \bibinfo{title}{The major trends shaping, breaking and consolidating
  global media by 2021}.
\newblock
  \bibinfo{howpublished}{\url{https://www.thedrum.com/news/2019/04/18/the-major-trends-shaping-breaking-and-consolidating-global-media-2021}}.
\newblock


\bibitem[\protect\citeauthoryear{Moody}{Moody}{2013}]%
        {declinedTraffic}
\bibfield{author}{\bibinfo{person}{Glyn Moody}.}
  \bibinfo{year}{2013}\natexlab{}.
\newblock \bibinfo{title}{Surprise: Paywalls Cause Massive Falls In Number Of
  Visitors - And Boost Competitors}.
\newblock
  \bibinfo{howpublished}{\url{https://www.techdirt.com/articles/20130920/09592024590/surprise-paywalls-cause-massive-falls-number-visitors-boost-competitors.shtml}}.
\newblock


\bibitem[\protect\citeauthoryear{Moses}{Moses}{2017}]%
        {wsJGoogleRef}
\bibfield{author}{\bibinfo{person}{Lucia Moses}.}
  \bibinfo{year}{2017}\natexlab{}.
\newblock \bibinfo{title}{The Wall Street Journal to close Google loophole
  entirely}.
\newblock
  \bibinfo{howpublished}{\url{https://digiday.com/media/wall-street-journal-close-google-loophole-entirely/}}.
\newblock


\bibitem[\protect\citeauthoryear{{Mozilla Foundation}}{{Mozilla
  Foundation}}{2019}]%
        {readability}
\bibfield{author}{\bibinfo{person}{{Mozilla Foundation}}.}
  \bibinfo{year}{2019}\natexlab{}.
\newblock \bibinfo{title}{{Readability.js}}.
\newblock \bibinfo{howpublished}{\url{https://github.com/mozilla/readability}}.
\newblock


\bibitem[\protect\citeauthoryear{Myllylahti}{Myllylahti}{2017}]%
        {myllylahti2017content}
\bibfield{author}{\bibinfo{person}{Merja Myllylahti}.}
  \bibinfo{year}{2017}\natexlab{}.
\newblock \showarticletitle{What Content is Worth Locking Behind a Paywall?
  Digital news commodification in leading Australasian financial newspapers}.
\newblock \bibinfo{journal}{\emph{Digital Journalism}} \bibinfo{volume}{5},
  \bibinfo{number}{4} (\bibinfo{year}{2017}), \bibinfo{pages}{460--471}.
\newblock


\bibitem[\protect\citeauthoryear{Nithyanand, Khattak, Javed, Vallina-Rodriguez,
  Falahrastegar, Powles, De~Cristofaro, Haddadi, and Murdoch}{Nithyanand
  et~al\mbox{.}}{2016}]%
        {nithyanand2016adblocking}
\bibfield{author}{\bibinfo{person}{Rishab Nithyanand},
  \bibinfo{person}{Sheharbano Khattak}, \bibinfo{person}{Mobin Javed},
  \bibinfo{person}{Narseo Vallina-Rodriguez}, \bibinfo{person}{Marjan
  Falahrastegar}, \bibinfo{person}{Julia~E Powles}, \bibinfo{person}{Emiliano
  De~Cristofaro}, \bibinfo{person}{Hamed Haddadi}, {and}
  \bibinfo{person}{Steven~J Murdoch}.} \bibinfo{year}{2016}\natexlab{}.
\newblock \showarticletitle{Adblocking and counter blocking: A slice of the
  arms race}. In \bibinfo{booktitle}{\emph{6th $\{$USENIX$\}$ Workshop on Free
  and Open Communications on the Internet ($\{$FOCI$\}$ 16)}}.
\newblock


\bibitem[\protect\citeauthoryear{Orem and Caio}{Orem and Caio}{2018}]%
        {bypassExt2}
\bibfield{author}{\bibinfo{person}{Rodrigo Orem} {and} \bibinfo{person}{Caio}.}
  \bibinfo{year}{2018}\natexlab{}.
\newblock \bibinfo{title}{Burlesco: Read news without subscribing, bypass the
  paywall}.
\newblock \bibinfo{howpublished}{\url{https://burles.co/en/}}.
\newblock


\bibitem[\protect\citeauthoryear{{Outline}}{{Outline}}{2018}]%
        {outline}
\bibfield{author}{\bibinfo{person}{{Outline}}.}
  \bibinfo{year}{2018}\natexlab{}.
\newblock \bibinfo{title}{Outline - Read \& annotate without distractions}.
\newblock \bibinfo{howpublished}{\url{https://outline.com/}}.
\newblock


\bibitem[\protect\citeauthoryear{Pachilakis, Papadopoulos, Markatos, and
  Kourtellis}{Pachilakis et~al\mbox{.}}{2019}]%
        {pachilakis2019no}
\bibfield{author}{\bibinfo{person}{Michalis Pachilakis},
  \bibinfo{person}{Panagiotis Papadopoulos}, \bibinfo{person}{Evangelos~P
  Markatos}, {and} \bibinfo{person}{Nicolas Kourtellis}.}
  \bibinfo{year}{2019}\natexlab{}.
\newblock \showarticletitle{No More Chasing Waterfalls: A Measurement Study of
  the Header Bidding Ad-Ecosystem}. In \bibinfo{booktitle}{\emph{Proceedings of
  the 19th Internet Measurement Conference}} \emph{(\bibinfo{series}{IMC
  2019})}.
\newblock


\bibitem[\protect\citeauthoryear{Papadopoulos, Diamantaris, Papadopoulos,
  Petsas, Ioannidis, and Markatos}{Papadopoulos et~al\mbox{.}}{2017a}]%
        {trackersWWW2016}
\bibfield{author}{\bibinfo{person}{Elias~P Papadopoulos},
  \bibinfo{person}{Michalis Diamantaris}, \bibinfo{person}{Panagiotis
  Papadopoulos}, \bibinfo{person}{Thanasis Petsas}, \bibinfo{person}{Sotiris
  Ioannidis}, {and} \bibinfo{person}{Evangelos~P Markatos}.}
  \bibinfo{year}{2017}\natexlab{a}.
\newblock \showarticletitle{The long-standing privacy debate: Mobile websites
  vs mobile apps}. In \bibinfo{booktitle}{\emph{Proceedings of the 26th
  International Conference on World Wide Web}}. International World Wide Web
  Conferences Steering Committee, \bibinfo{pages}{153--162}.
\newblock


\bibitem[\protect\citeauthoryear{Papadopoulos, Ilia, and Markatos}{Papadopoulos
  et~al\mbox{.}}{2018a}]%
        {truthMiners2018}
\bibfield{author}{\bibinfo{person}{Panagiotis Papadopoulos},
  \bibinfo{person}{Panagiotis Ilia}, {and} \bibinfo{person}{Evangelos~P
  Markatos}.} \bibinfo{year}{2018}\natexlab{a}.
\newblock \showarticletitle{Truth in Web Mining: Measuring the Profitability
  and Cost of Cryptominers as a Web Monetization Model}.
\newblock \bibinfo{journal}{\emph{arXiv preprint arXiv:1806.01994}}
  (\bibinfo{year}{2018}).
\newblock


\bibitem[\protect\citeauthoryear{Papadopoulos, Kourtellis, and
  Markatos}{Papadopoulos et~al\mbox{.}}{2019}]%
        {DBLP:journals/corr/abs-1805-10505}
\bibfield{author}{\bibinfo{person}{Panagiotis Papadopoulos},
  \bibinfo{person}{Nicolas Kourtellis}, {and} \bibinfo{person}{Evangelos
  Markatos}.} \bibinfo{year}{2019}\natexlab{}.
\newblock \showarticletitle{Cookie synchronization: Everything you always
  wanted to know but were afraid to ask}. In \bibinfo{booktitle}{\emph{The
  World Wide Web Conference}}. \bibinfo{pages}{1432--1442}.
\newblock


\bibitem[\protect\citeauthoryear{Papadopoulos, Kourtellis, and
  Markatos}{Papadopoulos et~al\mbox{.}}{2018b}]%
        {2018adcost}
\bibfield{author}{\bibinfo{person}{Panagiotis Papadopoulos},
  \bibinfo{person}{Nicolas Kourtellis}, {and} \bibinfo{person}{Evangelos~P
  Markatos}.} \bibinfo{year}{2018}\natexlab{b}.
\newblock \showarticletitle{The cost of digital advertisement: Comparing user
  and advertiser views}. In \bibinfo{booktitle}{\emph{Proceedings of the 2018
  World Wide Web Conference}}. International World Wide Web Conferences
  Steering Committee, \bibinfo{pages}{1479--1489}.
\newblock


\bibitem[\protect\citeauthoryear{Papadopoulos, Kourtellis, Rodriguez, and
  Laoutaris}{Papadopoulos et~al\mbox{.}}{2017b}]%
        {rtbPrices17}
\bibfield{author}{\bibinfo{person}{Panagiotis Papadopoulos},
  \bibinfo{person}{Nicolas Kourtellis}, \bibinfo{person}{Pablo~Rodriguez
  Rodriguez}, {and} \bibinfo{person}{Nikolaos Laoutaris}.}
  \bibinfo{year}{2017}\natexlab{b}.
\newblock \showarticletitle{If you are not paying for it, you are the product:
  How much do advertisers pay to reach you?}. In
  \bibinfo{booktitle}{\emph{Proceedings of the 2017 Internet Measurement
  Conference}}. ACM, \bibinfo{pages}{142--156}.
\newblock


\bibitem[\protect\citeauthoryear{Parr}{Parr}{2011}]%
        {NYTdecreasedTraffic1}
\bibfield{author}{\bibinfo{person}{Ben Parr}.} \bibinfo{year}{2011}\natexlab{}.
\newblock \bibinfo{title}{What Impact Has The New York Times Paywall Had on
  Traffic? [STATS]}.
\newblock
  \bibinfo{howpublished}{\url{https://mashable.com/2011/04/11/new-york-times-paywall-stats/}}.
\newblock


\bibitem[\protect\citeauthoryear{{Piano Inc.}}{{Piano Inc.}}{2015}]%
        {tinypass}
\bibfield{author}{\bibinfo{person}{{Piano Inc.}}}
  \bibinfo{year}{2015}\natexlab{}.
\newblock \bibinfo{title}{Overview - Tinypass for Developers}.
\newblock \bibinfo{howpublished}{\url{http://developer.tinypass.com/}}.
\newblock


\bibitem[\protect\citeauthoryear{{Piano Inc}}{{Piano Inc}}{2019}]%
        {algorithmicPaywall}
\bibfield{author}{\bibinfo{person}{{Piano Inc}}.}
  \bibinfo{year}{2019}\natexlab{}.
\newblock \bibinfo{title}{Algorithmic Paywall}.
\newblock
  \bibinfo{howpublished}{\url{https://docs.piano.io/algorithmic-paywall/}}.
\newblock


\bibitem[\protect\citeauthoryear{Ponsford}{Ponsford}{2017}]%
        {adOligopoly}
\bibfield{author}{\bibinfo{person}{Dominic Ponsford}.}
  \bibinfo{year}{2017}\natexlab{}.
\newblock \bibinfo{title}{Press Gazette launches Duopoly campaign to stop
  Google and Facebook destroying journalism}.
\newblock
  \bibinfo{howpublished}{\url{https://www.pressgazette.co.uk/press-gazette-launches-duopoly-campaign-to-stop-google-and-facebook-destroying-journalism/}}.
\newblock


\bibitem[\protect\citeauthoryear{{publicWWW}}{{publicWWW}}{2019}]%
        {publicWWW}
\bibfield{author}{\bibinfo{person}{{publicWWW}}.}
  \bibinfo{year}{2019}\natexlab{}.
\newblock \bibinfo{title}{Source Code Search Engine}.
\newblock \bibinfo{howpublished}{\url{https://publicwww.com/}}.
\newblock


\bibitem[\protect\citeauthoryear{Razaghpanah, Nithyanand, Vallina-Rodriguez,
  Sundaresan, Allman, and Gill}{Razaghpanah et~al\mbox{.}}{2018}]%
        {razaghpanah2018apps}
\bibfield{author}{\bibinfo{person}{Abbas Razaghpanah}, \bibinfo{person}{Rishab
  Nithyanand}, \bibinfo{person}{Narseo Vallina-Rodriguez},
  \bibinfo{person}{Srikanth Sundaresan}, \bibinfo{person}{Mark Allman}, {and}
  \bibinfo{person}{Christian Kreibich~Phillipa Gill}.}
  \bibinfo{year}{2018}\natexlab{}.
\newblock \showarticletitle{Apps, trackers, privacy, and regulators}. In
  \bibinfo{booktitle}{\emph{Proceedings of the Network and Distributed System
  Security Symposium}} \emph{(\bibinfo{series}{NDSS'18})}.
\newblock


\bibitem[\protect\citeauthoryear{Robles}{Robles}{2017}]%
        {paywallGoogle2}
\bibfield{author}{\bibinfo{person}{Patricio Robles}.}
  \bibinfo{year}{2017}\natexlab{}.
\newblock \bibinfo{title}{Google ditches first click free, embraces paywalls}.
\newblock
  \bibinfo{howpublished}{\url{https://econsultancy.com/google-ditches-first-click-free-embraces-paywalls/}}.
\newblock


\bibitem[\protect\citeauthoryear{Roettgers}{Roettgers}{2017}]%
        {paywallsDemocracy1}
\bibfield{author}{\bibinfo{person}{Janko Roettgers}.}
  \bibinfo{year}{2017}\natexlab{}.
\newblock \bibinfo{title}{BuzzFeed CEO Jonah Peretti: Paywalls Are Bad for
  Democracy}.
\newblock
  \bibinfo{howpublished}{\url{https://variety.com/2017/digital/news/buzzfeed-jonah-peretti-paywalls-democracy-1202593489/}}.
\newblock


\bibitem[\protect\citeauthoryear{Rutherford}{Rutherford}{2018}]%
        {paywallsGoogle}
\bibfield{author}{\bibinfo{person}{Sam Rutherford}.}
  \bibinfo{year}{2018}\natexlab{}.
\newblock \bibinfo{title}{Google Thinks It Can Make Paywalls Less of a Pain in
  the Ass}.
\newblock
  \bibinfo{howpublished}{\url{https://gizmodo.com/google-thinks-it-can-make-paywalls-less-of-a-pain-in-th-1823925612}}.
\newblock


\bibitem[\protect\citeauthoryear{Salmon}{Salmon}{2011}]%
        {NYTdecreasedTraffic2}
\bibfield{author}{\bibinfo{person}{Felix Salmon}.}
  \bibinfo{year}{2011}\natexlab{}.
\newblock \bibinfo{title}{The NYT paywall is working}.
\newblock
  \bibinfo{howpublished}{\url{http://blogs.reuters.com/felix-salmon/2011/07/26/the-nyt-paywall-is-working/}}.
\newblock


\bibitem[\protect\citeauthoryear{Saucier}{Saucier}{2018}]%
        {justRead}
\bibfield{author}{\bibinfo{person}{Zach Saucier}.}
  \bibinfo{year}{2018}\natexlab{}.
\newblock \bibinfo{title}{Just Read}.
\newblock \bibinfo{howpublished}{\url{https://justread.link/}}.
\newblock


\bibitem[\protect\citeauthoryear{{Scikit-learn developers}}{{Scikit-learn
  developers}}{2019}]%
        {sklearn}
\bibfield{author}{\bibinfo{person}{{Scikit-learn developers}}.}
  \bibinfo{year}{2019}\natexlab{}.
\newblock \bibinfo{title}{Scikit-learn: Machine Learning in Python}.
\newblock
  \bibinfo{howpublished}{\url{https://scikit-learn.org/stable/index.html}}.
\newblock


\bibitem[\protect\citeauthoryear{Simon and Graves}{Simon and Graves}{2019}]%
        {paywallNiemanlab}
\bibfield{author}{\bibinfo{person}{Felix Simon} {and} \bibinfo{person}{Lucas
  Graves}.} \bibinfo{year}{2019}\natexlab{}.
\newblock \bibinfo{title}{Across seven countries, the average price for
  paywalled news is about \$15.75/month}.
\newblock
  \bibinfo{howpublished}{\url{https://www.niemanlab.org/2019/05/across-seven-countries-the-average-price-for-paywalled-news-is-about-15-75-month/}}.
\newblock


\bibitem[\protect\citeauthoryear{Southern}{Southern}{2018}]%
        {mlpaywall}
\bibfield{author}{\bibinfo{person}{Lucinda Southern}.}
  \bibinfo{year}{2018}\natexlab{}.
\newblock \bibinfo{title}{How Swiss news publisher NZZ built a flexible paywall
  using machine learning}.
\newblock
  \bibinfo{howpublished}{\url{https://digiday.com/media/swiss-news-publisher-nzz-built-flexible-paywall-using-machine-learning/}}.
\newblock


\bibitem[\protect\citeauthoryear{Stewart}{Stewart}{2018}]%
        {adlergic}
\bibfield{author}{\bibinfo{person}{Duncan Stewart}.}
  \bibinfo{year}{2018}\natexlab{}.
\newblock \bibinfo{title}{Are Consumers 'Adlergic'? A Look at Ad-Blocking
  Habits}.
\newblock
  \bibinfo{howpublished}{\url{https://deloitte.wsj.com/cmo/2018/04/03/are-consumers-adlergic-a-look-at-ad-blocking-habits/}}.
\newblock


\bibitem[\protect\citeauthoryear{Tepper}{Tepper}{2017}]%
        {paywallsFacebook}
\bibfield{author}{\bibinfo{person}{Fitz Tepper}.}
  \bibinfo{year}{2017}\natexlab{}.
\newblock \bibinfo{title}{Facebook is now testing paywalls and subscriptions
  for Instant Articles}.
\newblock
  \bibinfo{howpublished}{\url{https://techcrunch.com/2017/10/19/facebook-is-now-testing-paywalls-and-subscriptions-for-instant-article/}}.
\newblock


\bibitem[\protect\citeauthoryear{Vallina-Rodriguez, Sundaresan, Razaghpanah,
  Nithyanand, Allman, Kreibich, and Gill}{Vallina-Rodriguez
  et~al\mbox{.}}{2016}]%
        {vallina2016tracking}
\bibfield{author}{\bibinfo{person}{Narseo Vallina-Rodriguez},
  \bibinfo{person}{Srikanth Sundaresan}, \bibinfo{person}{Abbas Razaghpanah},
  \bibinfo{person}{Rishab Nithyanand}, \bibinfo{person}{Mark Allman},
  \bibinfo{person}{Christian Kreibich}, {and} \bibinfo{person}{Phillipa Gill}.}
  \bibinfo{year}{2016}\natexlab{}.
\newblock \showarticletitle{Tracking the trackers: Towards understanding the
  mobile advertising and tracking ecosystem}.
\newblock \bibinfo{journal}{\emph{arXiv preprint arXiv:1609.07190}}
  (\bibinfo{year}{2016}).
\newblock


\bibitem[\protect\citeauthoryear{Vastel, Snyder, and Livshits}{Vastel
  et~al\mbox{.}}{2018}]%
        {vastel2018filters}
\bibfield{author}{\bibinfo{person}{Antoine Vastel}, \bibinfo{person}{Peter
  Snyder}, {and} \bibinfo{person}{Benjamin Livshits}.}
  \bibinfo{year}{2018}\natexlab{}.
\newblock \showarticletitle{Who filters the filters: Understanding the growth,
  usefulness and efficiency of crowdsourced ad blocking}.
\newblock \bibinfo{journal}{\emph{arXiv preprint arXiv:1810.09160}}
  (\bibinfo{year}{2018}).
\newblock


\bibitem[\protect\citeauthoryear{Wang}{Wang}{2018}]%
        {wsjPaywall}
\bibfield{author}{\bibinfo{person}{Shan Wang}.}
  \bibinfo{year}{2018}\natexlab{}.
\newblock \bibinfo{title}{After years of testing, The Wall Street Journal has
  built a paywall that bends to the individual reader}.
\newblock
  \bibinfo{howpublished}{\url{http://www.niemanlab.org/2018/02/after-years-of-testing-the-wall-street-journal-has-built-a-paywall-that-bends-to-the-individual-reader/}}.
\newblock


\bibitem[\protect\citeauthoryear{Weiner}{Weiner}{2007}]%
        {pocket}
\bibfield{author}{\bibinfo{person}{Nate Weiner}.}
  \bibinfo{year}{2007}\natexlab{}.
\newblock \bibinfo{title}{Pocket - Put knowledge in your Pocket}.
\newblock \bibinfo{howpublished}{\url{https://getpocket.com}}.
\newblock


\bibitem[\protect\citeauthoryear{{Wikimedia Foundation}}{{Wikimedia
  Foundation}}{2018}]%
        {wikipediaFunds2}
\bibfield{author}{\bibinfo{person}{{Wikimedia Foundation}}.}
  \bibinfo{year}{2018}\natexlab{}.
\newblock \bibinfo{title}{2016-2017 Fundraising Report}.
\newblock
  \bibinfo{howpublished}{\url{https://foundation.wikimedia.org/wiki/2016-2017_Fundraising_Report}}.
\newblock


\bibitem[\protect\citeauthoryear{Zarras, Kapravelos, Stringhini, Holz, Kruegel,
  and Vigna}{Zarras et~al\mbox{.}}{2014}]%
        {Zarras:2014:DAM:2663716.2663719}
\bibfield{author}{\bibinfo{person}{Apostolis Zarras},
  \bibinfo{person}{Alexandros Kapravelos}, \bibinfo{person}{Gianluca
  Stringhini}, \bibinfo{person}{Thorsten Holz}, \bibinfo{person}{Christopher
  Kruegel}, {and} \bibinfo{person}{Giovanni Vigna}.}
  \bibinfo{year}{2014}\natexlab{}.
\newblock \showarticletitle{The Dark Alleys of Madison Avenue: Understanding
  Malicious Advertisements}. In \bibinfo{booktitle}{\emph{Proceedings of the
  2014 Conference on Internet Measurement Conference}}
  \emph{(\bibinfo{series}{IMC '14})}.
\newblock


\end{thebibliography}
\end{document}